\documentclass[12pt, a4paper]{article}
\usepackage[scale={.75,.8}]{geometry}
\usepackage{pstricks}
\usepackage[latin2]{inputenc}
\usepackage{epic}
\usepackage{latexsym}
\usepackage{epsf}
\usepackage{epsfig}
\usepackage{amssymb,amsmath}
\usepackage{amsfonts}
\usepackage{dsfont}
\usepackage{graphicx}
\usepackage{subfigure}
\usepackage{multirow}
\def\beq{\begin{equation}}
\def\eeq{\end{equation}}
\newcommand{\bal}{\begin{align}}
\newcommand{\eal}{\end{align}}
\def\mtu{m_{\tilde{t}_1}}
\def\mtl{m_{\tilde{t}_2}}

\begin{document}






\title{\hspace{4.1in}\\\hspace{4.1in}{\small OUTP-0919P}\bigskip\\Naturalness and Focus Points with Non-Universal Gaugino Masses}
\author{D. Horton$^{1}$\thanks{email: \texttt{d.horton1@physics.ox.ac.uk}}~,
G. G. Ross$^{1}$\thanks{email: \texttt{g.ross1@physics.ox.ac.uk}}~\\$^{1}${\normalsize \textit{The Rudolf Peierls Centre for Theoretical
Physics,}}\\{\normalsize \textit{University of Oxford, 1 Keble Road, Oxford, OX1 3NP, UK}}\\}
\date{}
\maketitle

\begin{abstract}
Relations between the gaugino masses have been shown to alleviate the degree of fine-tuning in the MSSM. In this paper we consider specific models of supersymmetry breaking with gravity mediation and demonstrate that within both GUT and string constructions it is possible to generate these relations in a natural way. We have numerically studied the degree of fine-tuning in these models, including one-loop corrections, and have found regions of parameter space that can satisfy all known collider constraints with fine-tunings less than 20\%. We discuss some of the phenomenological features of these models within the regions of reduced fine-tuning.
\end{abstract}






\section{Introduction}\label{s.Intro}
Low energy supersymmetry \cite{c.LowEnergySUSY} (SUSY) is a well motivated proposal for physics beyond the standard model. In particular, the Minimal Supersymmetric Standard Model (MSSM) \cite{c.MSSM} predicts unification of the gauge couplings (to within a few percent) at a scale $M_X\sim 10^{16} \text{ GeV}$ and, together with R-parity conservation, provides a good dark matter candidate in the form of a neutralino. More significantly, perhaps, by stabilising the hierarchy between scales it explains why $M_Z\ll M_X$ and, through the soft breaking of supersymmetry, it may offer an explanation for the observed scale of electroweak symmetry breaking (EWSB). However, despite extensive searches performed at particle accelerators and dark matter detection experiments, no direct evidence for supersymmetry, or the Higgs, has yet been observed.

Following the absence of these signals at LEP, a lower bound of 114 GeV was placed on the mass of the Standard Model (SM) Higgs, at 95\% confidence \cite{c.LEPHiggsLimit}. Furthermore, limits have been set upon the masses of the supersymmetric particles, most notably the charginos, with\footnote{Where the precise limit depends upon the masses of the lightest neutralino, chargino and sleptons.} $m_{\chi}>88\text{---}102 \text{ GeV}$ \cite{c.LEPCharginoMassLimits}. Several studies have shown that these constraints, combined with assumptions regarding the mediation scale and the pattern of soft terms, imply that a significant fine-tuning of the MSSM parameters is required if the correct scale of EWSB is to be obtained \cite{c.FTPriceLEP.CEP,c.FTPriceLEP.BS,c.FTPriceLEP.CEOP,c.FTPriceLEP.KK}.

As is well known, it is the LEP limit on the SM Higgs mass that is the dominant source of this fine-tuning. Although the LEP limit does not strictly apply to the MSSM, it is relevant for regions of parameter space that contain a Higgs scalar with SM-like couplings. If CP-violating phases, $\theta_i$, are present in the soft terms, then this is not a generic feature of parameter space. However, the bounds on the electric dipole moments (EDMs) of the electron, neutron and muon place very strong constraints on these phases \cite{c.EDMs}. Without some mechanism to suppress the EDMs, these phases are required to take small values and in the remainder of this paper we shall restrict ourselves to the case $\theta_i\sim0$ or $\pi$. In the absence of these phases, a SM-like Higgs is somewhat difficult to avoid. In the decoupling limit, where the CP-odd Higgs has a mass $m_A\gg M_Z$, the lightest CP-even Higgs, $h^0$, is SM-like. Its mass, $M_{h^0}$, including the dominant one-loop corrections from stop-top loops, is given by:
\begin{align}\label{e.Higgs1loop}
M_{h^0}^2=M_Z^2\cos^{2}2\beta+\frac{3M_t^2h_t^2}{4\pi^2}\Big(\ln(\frac{M_{S}^2}{M_t^2})+\delta_t\Big)+\dotsb
\end{align}
which depends upon the top Yukawa coupling, $h_t$, the top mass, $M_t$ and the average stop mass $M_S^2=m_{\tilde{t}_1}m_{\tilde{t}_2}$ (where $m_{\tilde{t}_i}$ are the eigenvalues of the stop mass matrix). $M_{h^0}$ also depends upon a threshold correction, $\delta_t$, which is determined by the off-diagonal entry of the stop mass matrix $X_t=A_t-\mu\cot\beta$. From Eq~\eqref{e.Higgs1loop} it follows that, for $\cos^{2}2\beta\approx1$, the LEP limit requires a minimum stop mass:
\beq\label{e.StopBound}
M_S^2\gtrsim M_t^2\exp\bigg({\frac{2\pi^2}{3h_t^2}\frac{(M_{h^0}^2-M_Z^2)}{M_t^2}-\delta_t}\bigg)
\eeq
which depends sensitively upon both $h_t$ and $\delta_t$. Previous numerical studies \cite{c.Essig}, which include the two-loop $\mathcal{O}(\alpha_t^2, \alpha_t\alpha_s)$ corrections, have shown that for $M_t=173{ \text{GeV}}$ and $\delta_t\sim 0$ ($\delta_t\sim 1$) one typically requires $M_S\gtrsim 1 \text{ TeV}$ ($600 \text{ GeV}$). One can try to lower the bound on $M_S$ by generating a large threshold correction, although this comes at the cost of requiring large values for $A_t$, and generally saturates at values $\delta_t=\delta_t^{\rm Max}\sim \mathcal{O}(3)$. One can also consider the non-decoupling limit, in which $m_A\sim M_Z$, where the two CP-even Higgs scalars are comparable in mass and involve large mixtures of the gauge eigenstates. Through this mixing the lightest Higgs can have a suppressed coupling to the $Z$, thus permitting $M_{h^0}\lesssim 114 \text{ GeV}$ without conflict with LEP. The heavier Higgs, however, becomes SM-like in this case and, as pointed out by Derm\'{i}\~{s}ek and Gunion \cite{c.DermiGunion}, heavy stops are still required to lift this Higgs above the LEP bound. Hence, in the absence of CP-violating phases at least, the LEP bound imposes significant constraints on the stop sector.

In models where the supersymmetry breaking is mediated at a high scale $\Lambda\sim \mathcal{O}(M_X)$ these constraints have important implications for the scale of EWSB. Through radiative corrections the EWSB scale is related to the soft supersymmetry breaking terms \cite{c.IR} (our conventions for the soft supersymmetry breaking terms are described in Appendix A) and in general
\beq\label{e.VEVexpansion}
\lambda v^2=c^j m_j^2(\Lambda)+c^{jk}\mathcal{M}_j(\Lambda)\mathcal{M}_k(\Lambda)+c^{\mu k}\mu(\Lambda)\mathcal{M}_k(\Lambda)+c^{\mu}\mu^2(\Lambda)
\eeq
where $v^2=\langle H_u^0\rangle^2+\langle H_d^0\rangle^2$, whilst $m_i^2(\Lambda)$, $\mathcal{M}_i(\Lambda)$ and $\mu(\Lambda)$ are a scalar soft mass, dimension-1 soft term (either a gaugino mass, $M_i$, or trilinear term, $A_{i}$) and the supersymmetric Higgs mass, respectively, all defined at the mediation scale. The Higgs quartic coupling, $\lambda$, is given at tree-level by $\lambda^{(0)}=\tfrac{1}{4}(g^2+g_Y^2)$.
The dominant radiative corrections are again generated by the top-stop sector, with contributions to $v^2$ of the form
\beq\label{e.VEVstopcontribution}
\lambda v^2\supset\frac{3h_t^2}{16\pi^2}\ln\bigg(\frac{{\rm e}\Lambda^2}{m_{\tilde{t}_1}m_{\tilde{t}_2}}\bigg)\Big(m_{Q3}^2(\Lambda)+m_{U3}^2(\Lambda)+A_t^2(\Lambda)\Big)
\eeq
Because of the high mediation scale $\Lambda$ and $h_t\sim\mathcal{O}(1)$, the soft terms associated with the stop sector appear in Eq~\eqref{e.VEVexpansion} with $\mathcal{O}(1)$ coefficients. For the Snowmass point SPS 1a \cite{c.SPS}, with $\Lambda=M_X$, one finds
\beq\label{e.MZexpansion}
\frac{M_Z^2}{2}\approx \lambda v^2\approx 0.3 m_{Q3}^2(M_X)+0.3m_{U3}^2(M_X)+1.5 M_3^2(M_X)+0.1 A_t^2(M_X)-\mu^2(M_X)+\dotsb
\eeq
The problem of fine-tuning arises because the constraint given in Eq~\eqref{e.StopBound} requires soft terms that generate contributions to $M_Z$ that are far too large. As a result cancellations must occur between different soft terms, and this cancellation requires fine-tuning of the parameters at the mediation scale.

In order to measure the degree of fine-tuning we will follow previous studies of the MSSM \cite{c.BarbGiud,c.FTPriceLEP.BS,c.FTPriceLEP.CEOP} and take the fine-tuning at a point in parameter space to be $\Delta={\rm Max}\{|\Delta_i|\}$, where $\Delta_i$ is the fine-tuning with respect to a parameter $a_i$, defined as:
\beq\label{e.FTmeasure}
\Delta_i=\frac{\partial \ln M_Z}{\partial \ln a_i}
\eeq
Values of $\Delta\sim\mathcal{O}(1)$ indicate that the EWSB scale arises naturally in that region of parameter space, with no special tuning of the parameters required.

From the point of view that the MSSM is an effective theory below the mediation scale, each individual soft term is a parameter that should be included in the fine-tuning measure. Hence taking, for example, large scalar stop masses $m_{Q3}, m_{U3}\sim\mathcal{O}(1 \text{ TeV})$ to achieve $M_S\gtrsim 1 \text{ TeV}$ generates a fine-tuning $\Delta_{Q3}\sim 70\gg 1$. This large fine-tuning indicates that a cancellation must occur between different parameters. In this effective theory picture, where all soft masses are unrelated, such a cancellation would be unnatural. The probability $\rho$ of such a cancellation is often taken as $\rho=\Delta^{-1}\sim 1 \%$, which implies that the observed EWSB scale is a product of chance in this region of parameter space.

This fine-tuning may, however, be an artifact of the effective theory description. In a more complete theory of supersymmetry breaking the soft terms of the low energy Lagrangian will be related and, written in terms of the parameters of the true theory, the observed EWSB scale may prove to be natural with no fine-tuning at all. Previous studies of the MSSM have tried to identify relations that, if present amongst soft terms at the mediation scale, significantly reduce the level of fine-tuning \cite{c.FTPriceLEP.CEOP,c.FTPriceLEP.KK,c.NaturalnessfromStringRelations}. Most interestingly, some of these relations can be easily motivated within known models of supersymmetry breaking.

One such relationship is the universality of the scalar masses at the high scale:
\beq
m_i^2(M_X)=m_0^2
\eeq
which is a natural consequence of a flavour-blind mediation mechanism, such as supergravity. In this case the EWSB condition of Eq~\eqref{e.VEVexpansion} becomes:
\beq\label{e.ScalarFocusPoint}
\lambda v^2=c^m m_0^2 + c^{ij}\mathcal{M}_i(M_X)\mathcal{M}_j(M_X)+\dotsb
\eeq
It has been shown that, provided $\tan\beta\gtrsim\mathcal{O}(5)$, a cancellation occurs between the tree level soft masses and the radiative corrections, such that $c^m \ll 1$  \cite{c.Hyperbolic, c.Focus}. This reduces the dependence of $M_Z$ upon the parameter $m_0$, hence permitting $\Delta_{m_0}\sim 2c^m m_0^2/M_Z^2\lesssim 10$ even for large $m_0\sim 1 \text{ TeV}$. This permits the heavy stops required to satisfy the LEP bound on the Higgs without a large fine-tuning. 

Relations between gaugino masses have also been shown to reduce fine-tuning. In some of the earliest fine-tuning studies \cite{c.FTPriceLEP.KK} it was noted that $M_Z$ is strongly dependent upon the gluino mass $M_3(\Lambda)$, thanks to the large values of both $h_t$ and $\alpha_S$. Hence it was suggested that a small fine-tuning $|\Delta_{M_3}|<10$ requires a relatively light gluino, with $M_3(\Lambda)\lesssim 160 \text{ GeV}$, whilst the bino and wino masses may be much heavier. Later studies have since pointed out that relations between gaugino masses can permit a much heavier gluino without significant fine-tuning \cite{c.NaturalnessfromStringRelations}. The basic idea is that the gaugino masses are related to some common parameter, such that $M_i(\Lambda)=\eta_i M_{1/2}$, where the $\eta_i$ are fixed\footnote{By fixed, we mean that the $\eta_i$ are not continuous parameters. Instead they are discrete ratios, generated by some mechanism, and because of this they are not included in the fine-tuning measure.} ratios. In this case the contribution of the gaugino mass parameter $M_{1/2}$ to the EWSB scale is given by:
\beq
\lambda v^2=c^i m_i^2(\Lambda) + c^{M}M_{1/2}^2 + \dotsb
\eeq
and for some choice of the $\eta_i$ the co-efficient $c^M\ll 1$. Hence, one can have $\Delta_{M_{1/2}}\lesssim 10$ even with $M_{1/2}\sim\mathcal{O}(1 \text{ TeV})$. This idea was discussed recently by Abe, Kobayashi and Omura (AKO) \cite{c.AKO}, who identified, in a bottom-up approach, certain non-universal ratios which permit heavy gaugino masses without a large fine-tuning. Furthermore, they point out that these large gaugino masses can, through radiative corrections, generate the heavy stops needed to lift the Higgs above the LEP limit.

Of course the critical question is whether the gaugino mass ratios that reduce fine-tuning can be generated by a specific model of supersymmetry breaking, otherwise the initial choice of gaugino mass ratios is itself fine-tuned.
In this paper we consider in detail specific supersymmetry breaking models with gravity mediation that do reduce fine-tuning.
The layout of our paper is as follows. In Section~\ref{s.GauginoFP} we will identify a preferred set of gaugino mass ratios that will reduce fine-tuning. Some of this work will overlap with the study of AKO, but we include it here for completeness. In Section~\ref{s.Models} we will compare these ratios to those that are predicted by known mechanisms of generating non-universal gaugino masses in supergravity. For those models that predict ratios in the preferred range we have performed a detailed numerical study of the fine-tuning, which we present in Section~\ref{s.FT}. In Section~\ref{s.Pheno} we highlight some of the phenomenology of these models in regions of parameter space with $\Delta<10$. Finally, we conclude in Section~\ref{s.Conc}.

\section{Reduced fine-tuning through non-universal gaugino masses} \label{s.GauginoFP}

In this section we identify the relations between gaugino masses that reduce fine-tuning. We will work within the context of the R-parity conserving MSSM, and we will assume no additional field content below the unification scale $M_X$. We will assume that the supersymmetry breaking is mediated to the observable sector at a high scale $\Lambda$. In general $\Lambda$, and the field content between the scales $M_X$ and $\Lambda$ will depend upon the specific details of the model under consideration. In this paper we will make the simplifying assumption that $\Lambda=M_X$, and we will ignore the effects of threshold corrections due to states lying above the scale $M_X$. Furthermore we will assume that some mechanism, which we will discuss further in Section~\ref{s.Models}, generates gaugino masses that are related to a common parameter $M_{1/2}$:
\beq
M_i(M_X)=\eta_i M_{1/2}
\eeq
where the ratios $\eta_i$ are either fixed or discrete, i.e. cannot be continuously varied. Hence whilst the parameter $M_{1/2}$ must be included in the fine-tuning measure, the $\eta_i$ are not. Without loss of generality we can define $M_2(M_X)=M_{1/2}$ (i.e $\eta_2=1$) and we can always choose our phases such that $M_2(M_X)$ is positive. We will also assume that the scalar masses and trilinear terms adopt universal relations:
\begin{align}
m_i^2(M_X)&=m_0^2 & A_{ijk}(M_X)=A_0
\end{align}
Hence the continuous, dimension-1 parameters that must be included in the fine-tuning measure is the set $a=\{m_0, M_{1/2}, A_0, B(M_X), \mu(M_X)\}$. We will refer to a generic member of this set using a subscript, $a_i$.

In order to calculate the fine-tuning, it is necessary to determine the relationship between $v$ and the parameters $a_i$. After minimising the one-loop effective potential one obtains a relation between $v$ and the running soft terms of the form:
\beq\label{e.MZ}
\lambda^{(0)}v^2=-\frac{\tan^2\beta}{\tan^2\beta-1}~\overline{m}_{H_u}^2+\frac{1}{\tan^2\beta-1}~\overline{m}_{H_d}^2-|\mu|^2
\eeq
where $\lambda^{(0)}=\tfrac{1}{4}(g^2+g_Y^2)$ is the tree-level quartic coupling and $\overline{m}_{H_x}^2=m_{H_x}^2+\partial V^{(1)}/\partial H_x^2$. $V^{(1)}$ is the one-loop Coleman-Weinberg (CW) potential \cite{c.CW}. As usual, we define $\tan\beta=\langle H_u^0\rangle/\langle H_d^0\rangle$, which is also related to the soft terms:
\beq\label{e.tan}
\sin2\beta=\frac{2B\mu}{\overline{m}_{H_u}^2+\overline{m}_{H_d}^2+2|\mu|^2}.
\eeq

The dominant $\mathcal{O}(h_t^2)$ contribution to the CW potential is generated by the top-stop sector:
\begin{multline}\label{e.CWleading}
\frac{\partial V^{(1)}}{\partial H_u^2}=-\frac{3h_t^2}{16\pi^2}\Bigg((m_{Q3}^2+m_{U3}^2+X_t^2)\ln\bigg(\frac{{\rm e}Q^2}{m_{\tilde{t}_1}m_{\tilde{t}_2}}\bigg)\\
-\bigg[(\mtu^2-\mtl^2)+\frac{(\mtu^2+\mtl^2)}{(\mtu^2-\mtl^2)}X_t^2\bigg]\ln\bigg(\frac{\mtu}{\mtl}\bigg)-2h_t^2v^2\ln\bigg(\frac{m_{\tilde{t}_1}m_{\tilde{t}_2}}{M_t^2}\bigg)\Bigg)
\end{multline}
where $Q$ is the renormalization scale and the $m_{\tilde{t_i}}$ are the eigenvalues of the stop mass matrix. Note that all of the quantities appearing in Eqs~\eqref{e.CWleading} and \eqref{e.MZ} are running quantities, with an implicit dependence upon $Q$. In order to correctly resum large logarithms and minimise the corrections from the CW potential, it is evident from Eq~\eqref{e.CWleading} that one should choose a scale $Q\sim M_S= \sqrt{m_{\tilde{t_1}}m_{\tilde{t_2}}}$. In the remainder of this paper we will fix $Q=M_S$, unless stated otherwise. By solving the RGEs \cite{c.RGEs} the running soft terms at the scale $M_S$ can be related to the supersymmetry breaking parameters $a_i$:
\beq\label{e.RGEsolnsSOFT}
\begin{aligned}
m_i^2(M_S)&=z_i^{jk}a_ja_k  \\
&=z_i^mm_0^2+z_i^MM_{1/2}^2+z_i^AA_0^2+2z_i^{MA}M_{1/2}A_0 \\
\mathcal{M}_{\alpha}(M_S)&=\mathcal{Z}_{\alpha}^i a_i  \\
&=\mathcal{Z}_{\alpha}^{M}M_{1/2}+\mathcal{Z}_{\alpha}^A A_0 \\
B(M_S)&=B(M_X)+\mathcal{Z}_{B}^{M}M_{1/2}+\mathcal{Z}_{B}^A A_0
\end{aligned}
\eeq
whilst the $\mu$ parameter is given by:
\beq\label{e.RGEsolnMU}
\mu(M_S)=\mathcal{Z}_{\mu}\mu(M_X).
\eeq
The coefficients $z_i^{jk}$ and $\mathcal{Z}_{\alpha}^i$ have an implicit dependence upon the scale $Q=M_S$ and the boundary conditions at $M_X$. Using the expressions for the running quantities given in Eqs~\eqref{e.RGEsolnsSOFT} and \eqref{e.RGEsolnMU} one can then express Eq~\eqref{e.MZ} in terms of the parameters that describe the supersymmetry breaking. This gives a relation between these parameters and $v$:
\beq\label{e.VEVexpansion2}
\lambda v^2=c^m m_0^2+c^M M_{1/2}^2+c^{MA}M_{1/2}A_0+c^AA_0^2+c^\mu \mu^2+c^{M\mu}M_{1/2}\mu+c^{A\mu}A_0\mu
\eeq
where $\lambda=\lambda^{(0)}+\lambda^{(1)}$, and $\lambda^{(1)}$ is a one-loop contribution to the quartic coupling. It should be clear that the coefficients $c$ and $\lambda$, although dimensionless, depend upon the parameters $a_i$ and $v$ through the structure of the one-loop corrections and through the dependence of $\tan\beta$ upon the soft terms, which is given by Eq~\eqref{e.tan}.

Let us now consider the fine-tuning with respect to a parameter $a_i$. For the measure given in Eq~\eqref{e.FTmeasure} this is:
\beq\label{e.DeltaM}
\Delta_{i}=\frac{a_i^2}{M_Z^2}\frac{\partial M_Z^2}{\partial a_i^2}\approx2\lambda^{(0)} \frac{a_i^2}{M_Z^2} \frac{\partial v^2}{\partial a_i^2}
\eeq
where in the approximate equality we have neglected the self-energy of the Z boson, $\Pi_{ZZ}$, that contributes to its mass through $M_Z^2=2\lambda^{(0)}v^2+{\rm Re }~\Pi_{ZZ}$.
Using Eq~\eqref{e.VEVexpansion2} it is straightforward to calculate the derivative $\partial v^2/ \partial a_i^2$ and determine the fine-tuning. However, let us note that for the choice of renormalization scale $Q=M_S$ the contributions from the CW potential are suppressed by loop factors of $(4\pi)^2$ with no large logarithmic enhancement. Hence, to begin with, let us ignore these one-loop corrections and calculate the fine-tuning at the leading tree level. Furthermore, since we want to minimise the degree of fine-tuning we will restrict ourselves to the region of parameter space in which $\tan^2\beta\gg1$. As can be seen from Eq~\eqref{e.Higgs1loop} this maximises the tree level contribution to the Higgs mass, and thus alleviates the constraints on the stop sector that result in the fine-tuning problem. In this case it follows from Eq~\eqref{e.VEVexpansion2}
\begin{align}\label{e.MZlargeTan}
\lambda^{(0)}v^2&=-m_{H_u}^2(M_S)-|\mu|^2+\dotsb\nonumber \\
&=-z_{H_u}^mm_0^2-z_{H_u}^MM_{1/2}^2-z_{H_u}^AA_0^2-2z_{H_u}^{MA}M_{1/2}A_0-\mathcal{Z}_{\mu}^2\mu^2(\Lambda)+\dotsb
\end{align}
where the ellipses imply terms that are suppressed by factors of $\tan^2\beta$ or one-loop coefficients $(4\pi)^{2}$. The leading contribution to the  fine-tuning $\Delta_M$, due to the gaugino mass, is therefore given by:
\beq\label{e.DeltaMtree}
\Delta_{M}\approx-2\frac{z_{H_u}^MM_{1/2}^2+z_{H_u}^{MA}M_{1/2}A_0}{M_Z^2}
\eeq

It is clear from Eq~\eqref{e.DeltaMtree} that if the coefficients $z^M_{H_u}$ and $z^{MA}_{H_u}$ are suppressed, then one can have a small fine-tuning $|\Delta_M|<10$ even with a large gaugino mass $M_{1/2}$. For example, one can achieve $|\Delta_M|<10$ for $M_{1/2}=1 \text{ TeV}$ provided $z_{H_u}^M\lesssim0.04\sim (8\pi)^{-1}$, if $A_0=0$. The ratios $\eta_i$ that therefore reduce fine-tuning are those for which $z_{H_u}^M$ is suppressed.

In order to calculate the RG coefficients of Eq~\eqref{e.RGEsolnsSOFT} and their dependence upon the $\eta_i$ it is necessary to specify the boundary conditions at the unification scale. In particular, we need to know the value of $M_X$ together with the values of the gauge and Yukawa couplings at that scale. These couplings must be determined by matching them onto low energy data, including threshold corrections from superpartners. To perform this calculation we have used the software package \textsf{SOFTSUSY 2.0} \cite{c.SOFTSUSY}, together with the following low energy inputs: $\alpha_S(M_Z)^{\overline{MS}}=0.1176\pm0.002$, $\alpha^{-1}(M_Z)^{\overline{MS}}=127.918$, $M_Z=91.1876 \text{ GeV}$, $m_b(m_b)^{\overline{MS}}=4.20 \text{ GeV}$ \cite{c.PDG} and $M_t=172.6\pm1.4 \text{ GeV}$ \cite{c.TopQuarkMass}. Shown in Table~\ref{t.Couplings} are the high scale couplings obtained for a number of Snowmass benchmark points \cite{c.SPS}. Note that the errors in $M_X$, $h_t(M_X)$ and $g_3(M_X)$ are the propagated errors due to the $1\sigma$ uncertainties in the top pole mass $M_t$ and $\alpha_S(M_Z)$, added in quadrature.

\begin{table}[bt]
\begin{center}
\begin{tabular}{|c|c|c|c|c|}
\hline
Benchmark Point&$M_X$ / $10^{16}$ GeV&$h_t(M_X)$&$g_3(M_X)$&$g_1(M_X)$\\
\hline
1a & 2.39(2) & 0.494(18) & 0.708(2) & 0.721\\
2 & 2.38(2) & 0.506(18) & 0.703(2) & 0.714 \\
3 & 1.97(2) & 0.498(18) & 0.705(2) & 0.716\\
5 & 2.15(2) & 0.508(20) & 0.707(2) & 0.718 \\
\hline
\end{tabular}
\end{center}
\caption{\small Values for $M_X$ and the couplings calculated using SOFTSUSY for four different Snowmass points. The uncertainty is calculated by allowing $M_t$ and $\alpha_S(M_Z)$ vary over their one $\sigma$ range, and then combining the resulting uncertainties in quadrature. The uncertainty in $g_1$ is small, typically less then $0.1\%$.}
\label{t.Couplings}
\end{table}

Because the couplings depend upon threshold corrections they are functions of the spectrum of superpartners, and hence of the parameters $a_i$ and $\eta_i$. However, we note from Table~\ref{t.Couplings} that the uncertainty due to these unknown threshold corrections is relatively small; comparable to the uncertainty due to errors in the low energy data. Thus, in what follows, we will fix these high scale couplings to the values appropriate for one supersymmetric spectrum, and then neglect further threshold dependence as we adjust the ratios $\eta_{i}$ and the parameters $a_i$.

Using the central values of SPS point 1a we find that the coefficients $z^{M}_{H_u}$ and $z^{MA}_{H_u}$ have the following form at a scale $Q=500\text{ GeV}$:
\begin{align}
z^M_{H_u}&= -1.53 \eta_3^2+0.21 -0.13\eta_3-0.020\eta_1\eta_3+0.007\eta_1^2-0.005\eta_1\\
z^{MA}_{H_u}&=0.29\eta_3+0.07+0.01\eta_1
\end{align}
It is clear that the contributions from the gluino mass are the most significant, as they are generated by the larger, and faster running, couplings $h_t$ and $\alpha_3$. Note also that, in $z^M_{H_u}$, the coefficient of $\eta_3^2$ is negative because the gluino mass feeds into the evolution of $m_{H_u}^2$ through a Yukawa coupling. This is critically important, since it permits a set of ratios $\eta_i$ such that $z^M_{H_u}(Q=500\text{ GeV})=0$. We note that there is no set of ratios such that both $z^M_{H_u}=z^{MA}_{H_u}=0$. For $|\eta_1|\lesssim\mathcal{O}(1)$ we observe that the coefficient $z^M_{H_u}(Q=500 \text{ GeV})$ vanishes for $\eta_3\approx0.33, -0.42$ , whilst $z^{MA}_{H_u}\approx0.17, -0.05$ respectively.

\begin{figure}[t]
\begin{center}
\includegraphics[width=\textwidth]{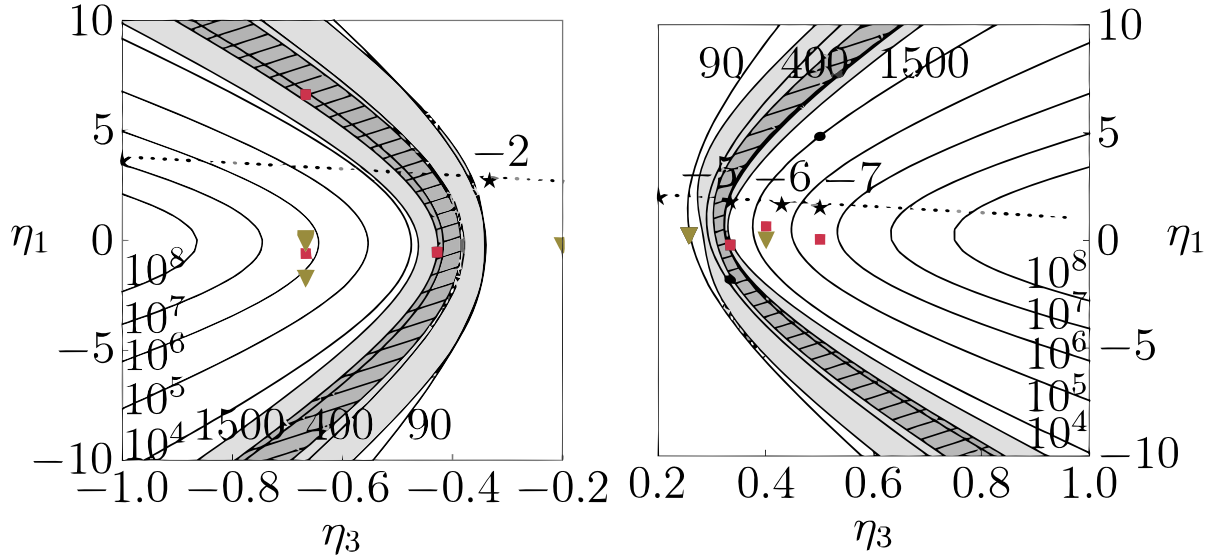}
\end{center}
\caption{\small A contour plot of the gaugino focus point scale $Q^{M} \text{ (GeV)}$, as a function of the gaugino mass ratios $\eta_1$ and $\eta_3$. The light grey (dark grey, hatched) region indicates the ratios that will permit $\Delta_{M}\lesssim10$ for $M_S=600 \text{ GeV}$ ($M_S=1\text{ TeV}$). The filled circles, squares and triangles indicate the mass ratios predicted by the $SU(5)$, $SO(10)$ and $E_6$ GUT models, respectively, that are discussed in Section~\ref{s.Models}. The stars give the prediction of a string model with moduli-dominated SUSY breaking (the O-II model \cite{c.TheoryOfSoftTerms}), for various integer values of the parameter $\delta_{GS}$. The ratios that are permitted within models with mixed Moduli-Anomaly breaking lie upon the dashed line.}
\label{f.QM}
\end{figure}

From this discussion it should be clear that for any choice of ratios $\eta_i$ the coefficient $z^M_{H_u}(Q)$ will vanish at some scale. At this scale the running soft mass $m_{H_u}^2$ will become independent of the parameter $M_{1/2}$, provided $A_0$ vanishes. This is analogous to the scalar focus point scale \cite{c.Focus}, where $m_{H_u}^2$ becomes independent of the scalar mass parameter $m_0$. 
From the perspective of an RG analysis it is this scalar focus point, and its proximity to the TeV scale, which is responsible for the smallness of the coefficient $c^m$ appearing in Eq~\eqref{e.ScalarFocusPoint}. 
Because of this analogy we will call the scale at which $z^M_{H_u}=0$ the gaugino focus point\footnote{This ignores the minor technical point that the gaugino focus point involves cancellations between radiative corrections, whilst the scalar focus point involves cancellations between tree level and radiative pieces.}, $Q^M$. In Fig~\ref{f.QM} we plot this focus point scale in the $\eta_3$---$\eta_1$ plane.

For $M_S\approx Q^M$ the coefficient $z^M_{H_u}(Q=M_S)$, that appears in the fine-tuning $\Delta_M$, has the following form:
\beq\label{e.zMexpansion}
z^M_{H_u}(M_S)\approx-\frac{3}{8\pi^2}\bigg(h_t^2(z_{Q3}^M+z_{U3}^M+(\mathcal{Z}_{A_t}^M)^2)-g_2^2(\mathcal{Z}_2^M)^2-\frac{1}{5}g_1^2(\mathcal{Z}_1^M)^2\bigg)\bigg|_{Q=Q^M}\ln\big(\frac{Q^M}{M_S}\big)
\eeq
where the coefficients $z_i^M$ and $\mathcal{Z}_{\alpha}^M$ that appear in Eq~\eqref{e.zMexpansion} are given in Appendix B. Thus, provided $Q^M$ is sufficiently close to $M_S$ the coefficient $z_{H_u}^M$ can be suppressed so as to permit large values of $M_{1/2}$ without a significant fine-tuning. Because the coefficient $z^{MA}_{H_u}$ cannot be simultaneously small, $|\Delta_M|<10$ will require that $A_0$ is small for large $M_{1/2}$. Hence we will restrict ourselves to the plane $A_0=0$ for the remainder of this paper\footnote{Abe et. al. have explored the possibility of allowing a large $A_0$ by postulating a further relation of the form $A_0=\eta_0M_{1/2}$.}.

As we discussed in Section~\ref{s.Intro} the LEP bound on the Higgs places a lower bound on the stop mass $M_S$ through Eq~\eqref{e.StopBound}. For $m_0=A_0=0$, one can determine the value of $M_{1/2}=\overline{M}$ which achieves this minimum stop mass for a given set of ratios $\eta_i$. This is straightforward to calculate using the RG coefficients given in Appendix B. Moreover, $|\Delta_{\mu}|=|1-\Delta_{M}|<|\Delta_M|$ in this tree-level approximation, thus $\Delta=|\Delta_M|$. The fine-tuning introduced by the LEP bound is then, for a particular set of ratios $\eta_i$,
\beq\label{e.DeltaMEst}
\Delta\approx \bigg|-2 \frac{z_{H_u}^M(M_S)\overline{M}^2}{M_Z^2}\bigg|
\eeq
In Fig~\ref{f.QM} we have highlighted, in the shaded and hatched regions, the ratios that have $\Delta<10$ for $M_S= 600 \text{ GeV}$ and $M_S=1 \text{ TeV}$, respectively. These values of $M_S$, according to a previous numerical study \cite{c.Essig}, are sufficient to lift the Higgs above the LEP bound when $\delta_t\sim 1$ and $0$, respectively. The ratios in these shaded regions are therefore the `preferred' ratios that reduce fine-tuning. An interesting question to ask is whether any known models {\it predict} such ratios, and we will turn to this in Section~\ref{s.Models}.

It is worth making a number of comments about this analysis. We have investigated the effect of uncertainties in the low energy data, namely in $M_t$ and $\alpha_S(M_Z)$, combined with uncertainties associated with the unknown threshold corrections. Allowing the low energy variables to shift by their $1\sigma$ errors and also choosing different Snowmass benchmark points, we observe only minor shifts ($\lesssim 1 \%$) in the position of the preferred region. What is more important is the effect of calculating $\Delta$ beyond the tree level approximation. For the ratios with a focus point $Q^M\sim M_S$ it is clear that $z_{H_u}^M$ will be comparable to one-loop contributions. Therefore, in the preferred region of Fig~\ref{f.QM}, one expects that the fine-tuning given by Eq~\eqref{e.DeltaMEst} is only an order of magnitude estimate. Thus the results presented in Fig~\ref{f.QM} should be used only as a guide to which ratios deserve a more detailed study. To determine the fine-tuning more accurately it will be necessary to calculate all of the one-loop and $\tan\beta$ suppressed contributions. Furthermore, the RGEs should be solved at the two-loop level, since the sub-leading logarithms arising at two loops are comparable to the one-loop contributions from the CW potential. We will perform such a study, numerically, for any model that predicts ratios in the preferred region.

\section{Non-universal gaugino masses from GUTs and strings}\label{s.Models}
In the previous section we have identified a preferred set of gaugino mass ratios that, potentially, could alleviate fine-tuning within the MSSM. Hence it is clearly of interest to examine whether any known models of supersymmetry breaking predict these ratios. For simplicity, we will restrict ourselves to models where the supersymmetry breaking is mediated by gravity alone. In this case there are several known mechanisms for generating non-universal gaugino masses, which we will briefly review here.

In general, the gaugino masses for a supergravity model are given by \cite{c.Cremmer}:
\beq\label{e.SUGRAGauginoMass}
M_{\alpha\beta}=\frac{1}{2{\rm Re} f_{\alpha\beta}}e^{-G/2}G^{i}(G^{-1})^j_i(\partial f^\ast_{\alpha\beta}/\partial {\phi^{j}}^\ast)
\eeq
where $G$ is the K\"ahler potential, a real gauge singlet function of the chiral superfields $\Phi_i$, and $f_{\alpha\beta}$ is the gauge kinetic function, an analytic function of the $\Phi_i$ that transforms under the gauge group as the symmetric product of two adjoints, $(Adj\times Adj)_S$. From \eqref{e.SUGRAGauginoMass} it is clear that non-universal gaugino masses will result if $f_{\alpha\beta}$ has some non-trivial gauge structure.

In a GUT this structure can be generated by a Higgs field, $\Sigma$, provided it transforms under the gauge group $G$ in an irreducible representation $R\subset (Adj\times Adj)_S$. If this field takes a vacuum expectation value (VEV), spontaneously breaking $G$ into a subgroup\footnote{Where this subgroup must include the Standard Model.} $H$, it can produce a gauge non-singlet contribution to $f_{\alpha\beta}$. In particular, for a gauge kinetic function of the form:
\begin{align}
f_{\alpha\beta}&=\delta_{\alpha\beta}+B(\Phi_i)\frac{\Sigma_{\alpha\beta}}{M_P}
\end{align}
where $B$ is an arbitrary (gauge singlet) function of the $\Phi_i$, the gaugino mass matrix is given by \cite{c.NonUniGauginos}:
\beq
M_{\alpha\beta}={\eta}_{\alpha}\delta_{\alpha\beta}M\langle\Sigma\rangle
\eeq
where the discrete coefficients $\eta_{\alpha}$ are determined by $R$ and $H$. For the GUT groups $SU(5)$, $SO(10)$ and $E_6$ the possible ratios have been determined and can be found in the literature\footnote{The recent study by Martin \cite{c.MartinNonUni} has highlighted possible errors in the previous study of $SO(10)$ \cite{c.SO10GauginoMasses}. In this work we use the results of Martin.} \cite{c.NonUniGauginos,c.SO10GauginoMasses,c.MartinNonUni}.

In Fig \ref{f.QM} we have plotted the focus point scale for the ratios $\eta_{\alpha}$ predicted by the GUT models. From the figure we note that there are several models that have $Q^M$ in the region that may permit low fine-tuning. The relevant details of these models are given in Table~\ref{t.GUTratios}, and we will study these in more detail in Section~\ref{s.FT}. As pointed out by Martin \cite{c.MartinNonUni}, the 54 breaks $SO(10)$ to a subgroup, $SU(2)\times SO(7)$, which cannot be reconciled with the chiral fermion content of the SM. Hence the 54 is not a realistic model. However, it does generate mass ratios that illustrate an interesting phenomenological structure and, pending the identification of a viable model with mass ratios in this region, we will include it in our phenomenological survey.

\begin{table}[tbp]
\begin{center}
\begin{tabular}{|c|c|c|c|c|}
\hline
$G$&$R$&$H$&$\eta_3$&$\eta_1$\\
\hline
\multirow{1}{*}{$SU(5)$}&75&\multirow{1}{*}{$SU(3)\times SU(2) \times U(1)$}&1/3&-5/3\\
\hline
\multirow{3}{*}{$SO(10)$}&$54$& $SU(2)\times SO(7)$ &$-3/7$&$-3/7$\\
&210&$[SU(5)'\times U(1)]_{\rm flipped}$&$1/3$&$-1/15$\\
&770&$[SU(5)'\times U(1)]_{\rm flipped}$& $-2/3$&$101/15$\\

\hline
\end{tabular}
\end{center}
\caption{\small The ratios of gaugino masses generated at $M_X$ for the subset of $SU(5)$, $SO(10)$ and $E_6$ GUTs \cite{c.NonUniGauginos,c.SO10GauginoMasses,c.MartinNonUni} that appear in the preferred, shaded region of Fig~\ref{f.QM}.}
\label{t.GUTratios}
\end{table}

String theories also provide a mechanism for generating non-trivial structure in $f_{\alpha\beta}$, in the form of contributions related to anomalies and string threshold corrections \cite{c.TheoryOfSoftTerms,c.StringSoft}. Although these arise as one loop corrections to the gauge kinetic function, they can generate large non-universality in the gaugino masses provided that the supersymmetry breaking occurs in the moduli, and not the dilaton, fields. The form of the string threshold corrections is highly model dependent, as they are sensitive to a number of factors, including: the choice of manifold for the compactified dimensions, the modular weights of the observable sector fields and the moduli VEVs. It is well beyond the scope of this work to survey all the possibilities; here we will only consider a simplified model based on an orbifold compactification. This model, coined the `O-II' model \cite{c.TheoryOfSoftTerms}, generates ratios\footnote{Note that, unlike the case explicitly considered in \cite{c.TheoryOfSoftTerms}, we assume VEVs for the moduli fields that are far from the self-dual point. In this case the contribution from the superconformal anomaly can be neglected \cite{c.StringSoft} and a large hierarchy between the gaugino and scalar masses can be avoided.} $\eta_{\alpha}\propto(-\delta_{GS}+b_{\alpha})$, where $b_{\alpha}=(33/5, 1, -3)$ is the one-loop beta function coefficient for the gauge couplings, whilst $\delta_{GS}$ is a constant\footnote{This constant is associated with the Green-Schwarz counter term that must be introduced to cancel anomalies.} that is required to be a negative integer by an anomaly cancellation condition. Since $\delta_{GS}$ is a discrete parameter it need not be included in our measure of fine-tuning.

Non-universal gaugino masses can also be generated in the so-called `Mixed Moduli-Anomaly' (MMA) scenario \cite{c.MMA}, where the moduli fields and  superconformal anomaly generate comparable contributions to the soft terms. In such models the gaugino mass ratios have the same structure as in the `O-II' model: $\eta_{\alpha}=a+b_{\alpha}$, where $a$ is the ratio of the moduli and anomaly contributions. In this case the parameter $a$ is related to topological quantities, and is a rational number involving the ratios of fluxes. Despite the fact $a$ takes only discrete values, if the interval between these values are small, it should be included in the fine-tuning measure. In this case there is no reduction in the fine-tuning. This should be compared with
the O-II model where the Green-Schwarz term takes on only integer values and thus does not contribute to the fine-tuning measure.

The mass ratios predicted by the MMA scenario lie along the dashed line shown in Fig~\ref{f.QM}, whilst those of the O-II model are indicated by stars. We note that the O-II with $\delta_{GS}=-5$, which has gaugino mass ratios $\{\eta_1=29/15, \eta_3=1/3\}$, possesses a set of ratios that may permit low fine-tuning. In fact, this point produces the same gaugino mass ratios predicted by the `Mirage Mediation' point of the MMA scenario, which has previously been demonstrated to possess low fine-tuning \cite{c.MMA}. This set of ratios is characterised by the unification of the gaugino masses at a low scale. Because the MMA scenario is complicated by the need to assign modular weights to each of the fields, we will postpone its discussion to a later paper. We simply note here that a low scale gaugino focus point can, for an appropriate choice of $a$, be achieved in the MMA with $M_3M_2<0$.

In order to make more quantitative statements about the naturalness of these models we must calculate the fine-tuning beyond the tree level approximation. We have done this numerically and we discuss our results in the following section. Note that, henceforth, we will refer to the O-II model with $\delta_{GS}=-5$ as simply the `O-II', and we will refer to the GUT models by the representation $R$ that is listed in Table~\ref{t.GUTratios}.

\section{Fine-tuning in the O-II and GUT models}\label{s.FT}
Using SOFTSUSY, together with the low-energy inputs discussed in Section \ref{s.GauginoFP}, we have calculated the fine-tuning in a fixed grid scan over the parameters $m_0$ and $M_{1/2}$. The derivative $\partial M_Z / \partial a_i$ is calculated by SOFTSUSY using the following procedure:
 \begin{itemize}
\item A parameter $a_i$ is shifted by a small quantity at the scale $M_X$.
\item The couplings and soft-masses are run down to the low scale $M_S$ using two loop RGEs.
\item The one-loop potential is then minimized to obtain the new Higgs VEV,  $v$.
\item This procedure is then repeated, calculating $v$ for increasingly smaller shifts of $a_i$.
\end{itemize}
By default, this procedure neglects the $v$ dependence of masses (of the fermions and scalars), appearing in the one-loop potential. In order to calculate the fine-tuning correctly at the one-loop level we modified this procedure, so that the one-loop potential is correctly minimised including the dependence of all masses upon $v$.

\begin{figure}[!t]
  \centering
     \subfigure[54]{
          \label{f.54FT}
          \includegraphics[width=0.45\textwidth]{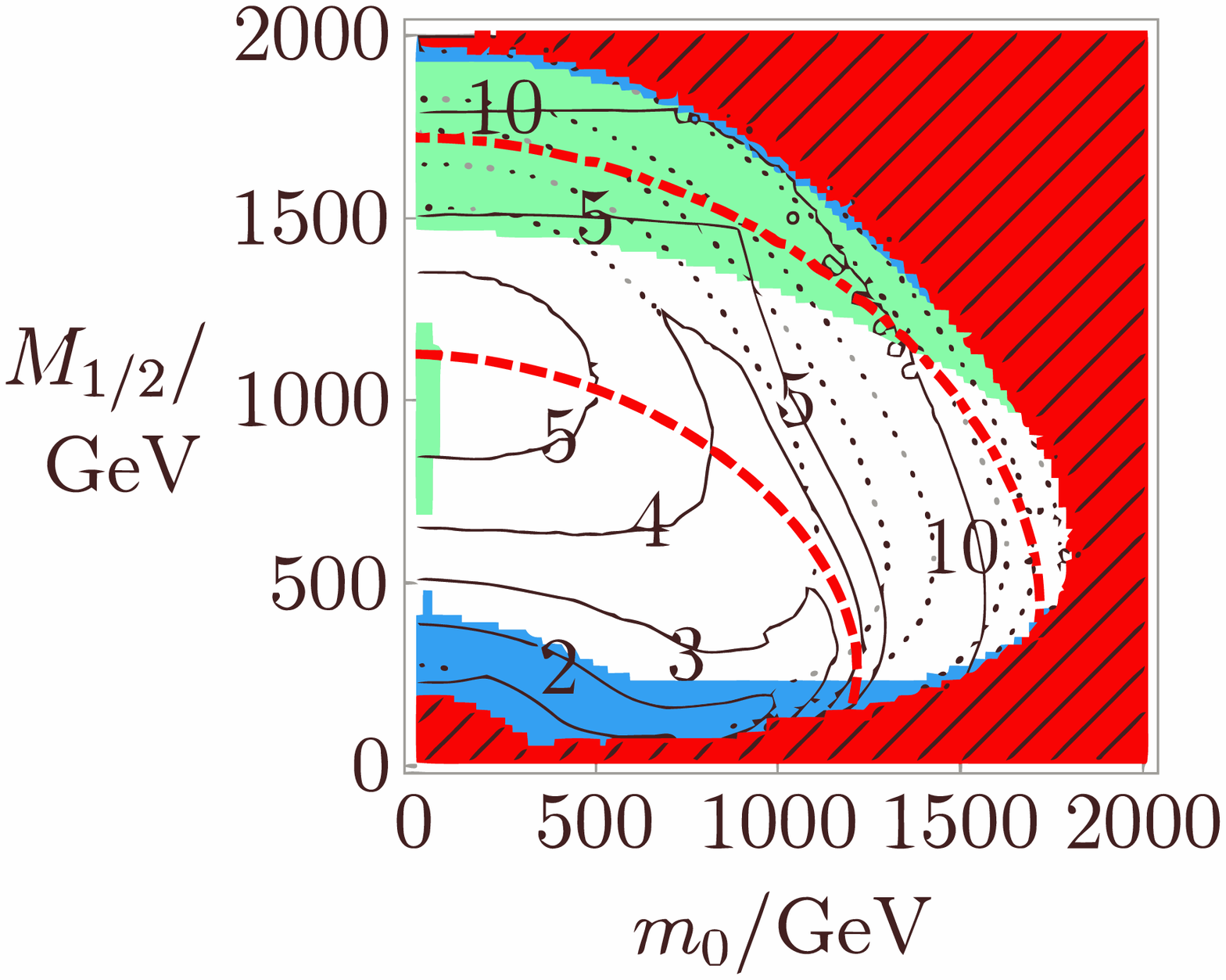}
          }
     \subfigure[210]{
          \label{f.210FT}
          \includegraphics[width=0.45\textwidth]{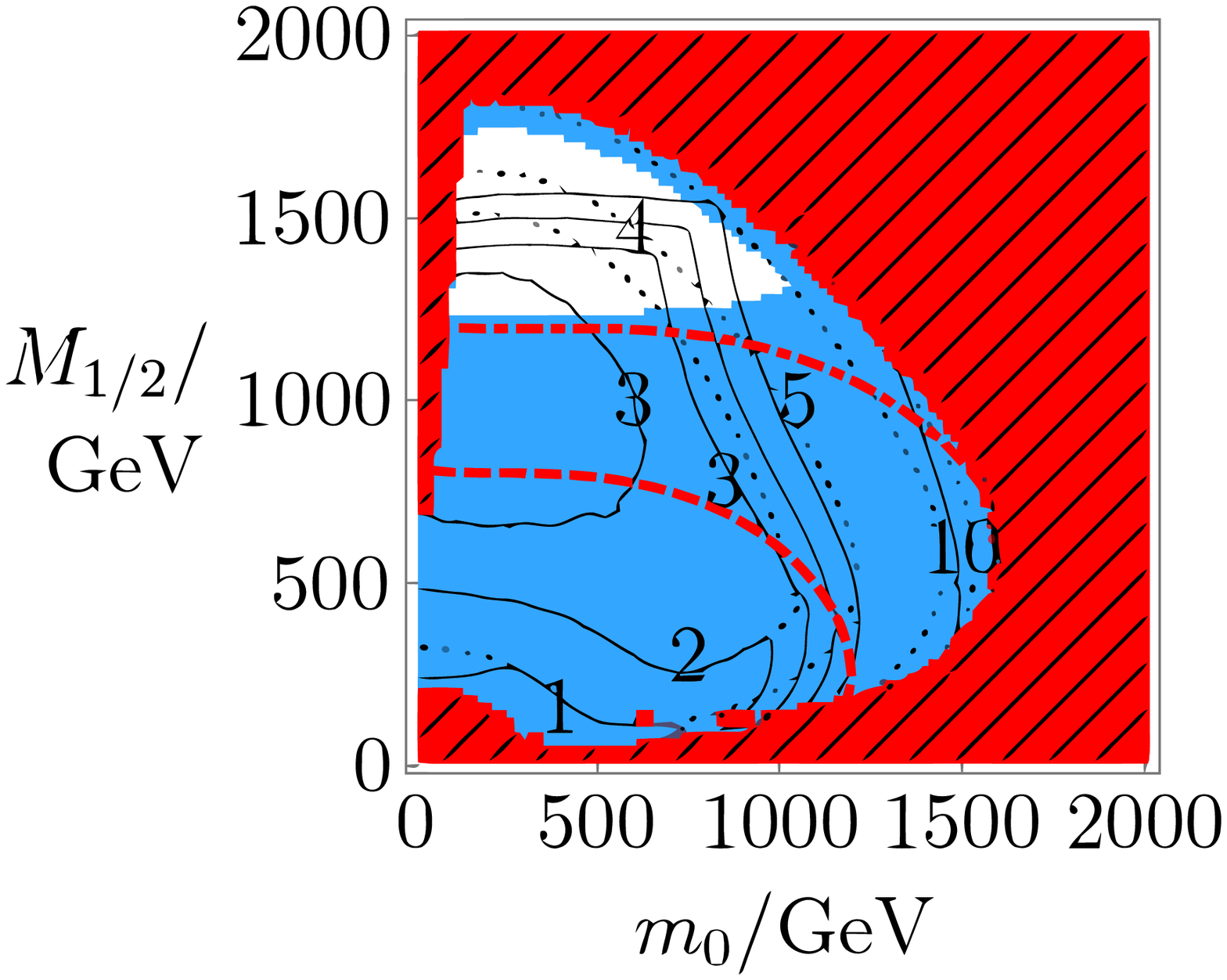}
          }
     \subfigure[770]{
           \label{f.770FT}
           \includegraphics[width=0.45\textwidth]{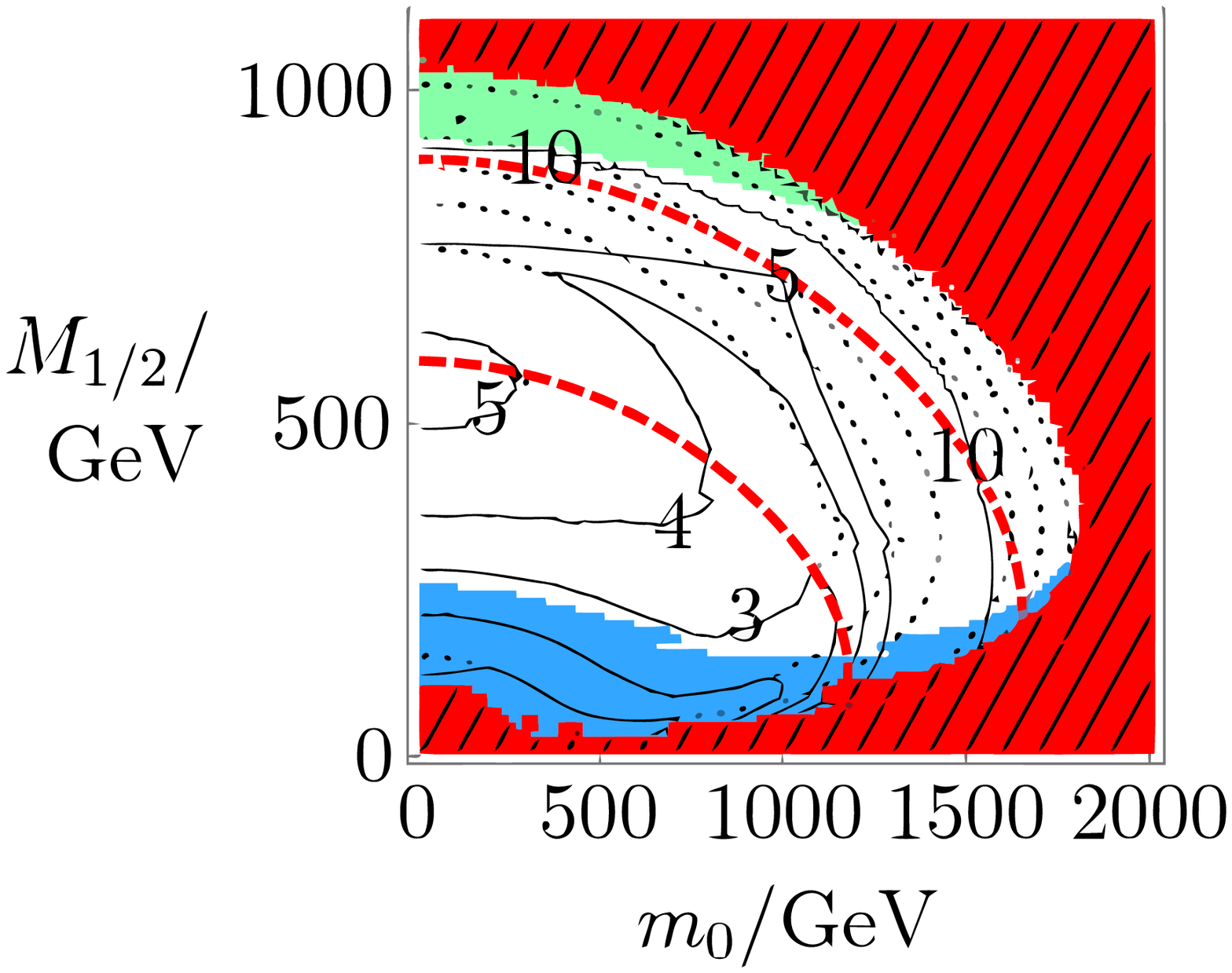}
           }
     \subfigure[O-II]{
           \label{f.OIIFT}
           \includegraphics[width=0.45\textwidth]{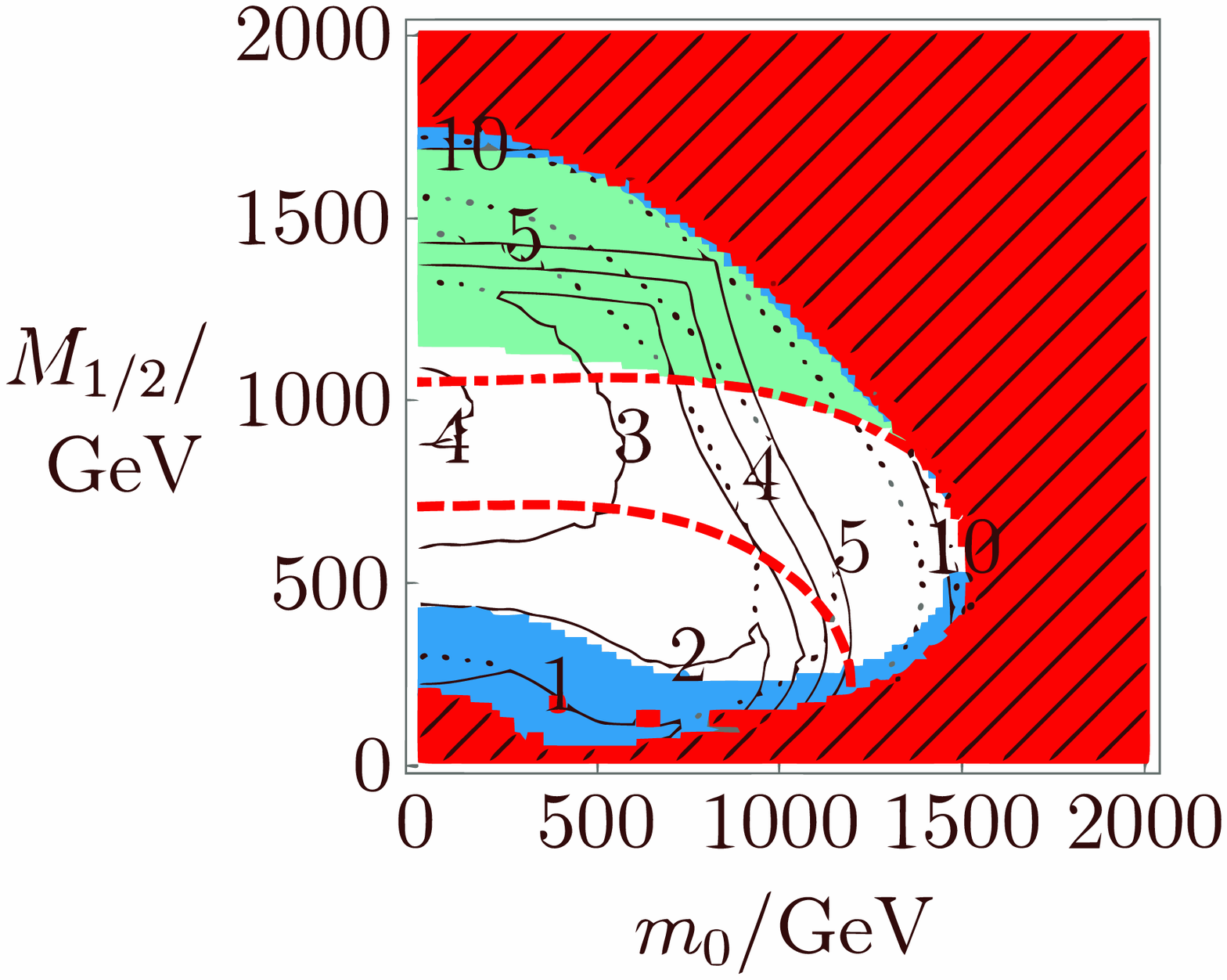}
           }
     \caption{\small The solid (black) contours indicate the fine-tuning in the $m_0$---$M_{1/2}$ plane. This is for the hypersurface in parameter space with $\tan\beta=10$, $A_0=0$ and $\mu>0$. The (red) dashed and (red) dot-dashed contours indicate where the Higgs mass, $m_{h^0}$, is 111 GeV and 114 GeV, respectively. The dotted contours are lines of constant $\Delta_{\mu}$.}
     \label{f.FT}
\end{figure}

\begin{figure}[!t]
  \centering
     \subfigure[54]{
          \label{f.54mFT}
          \includegraphics[width=0.45\textwidth]{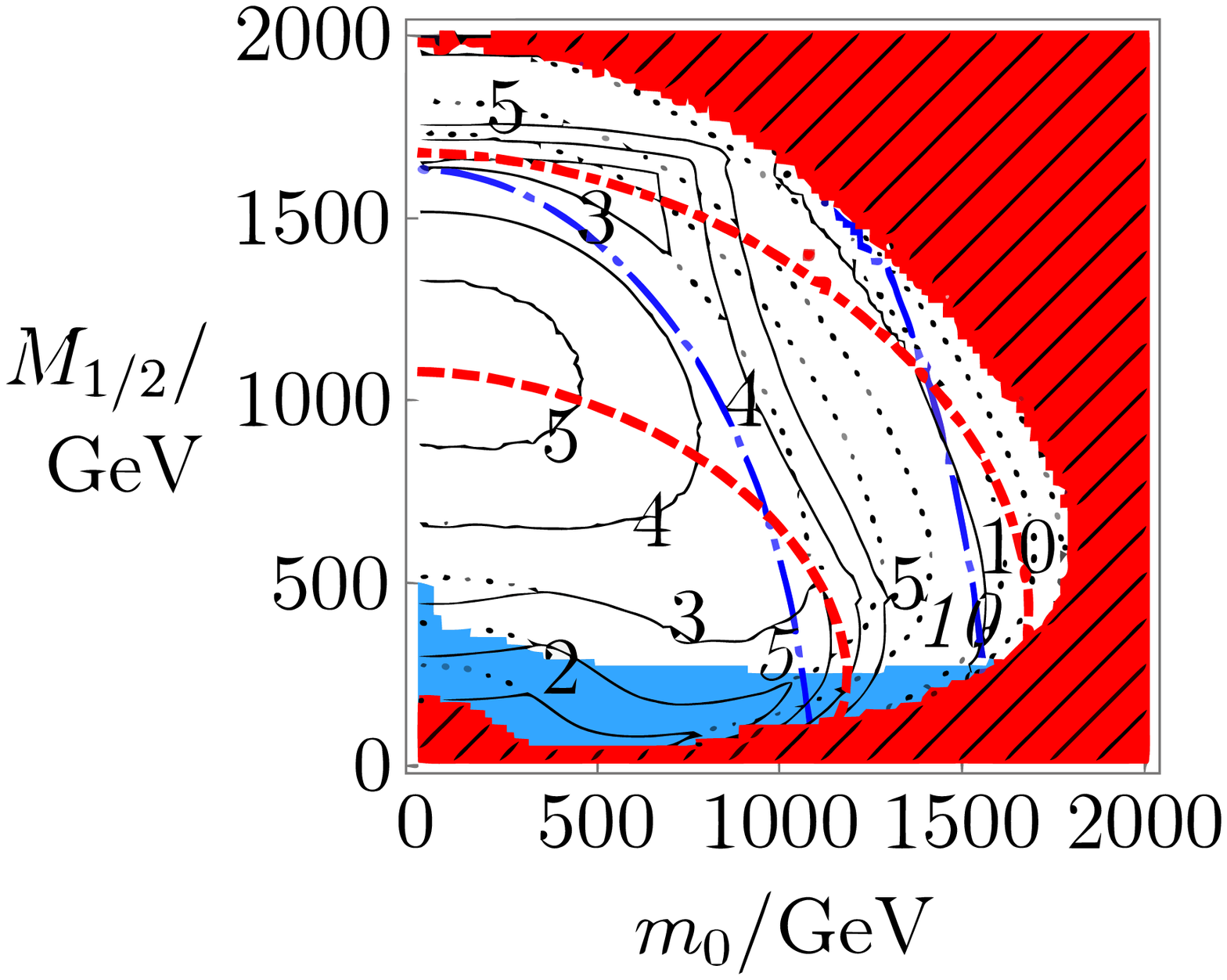}
          }
     \subfigure[210]{
          \label{f.210mFT}
          \includegraphics[width=0.45\textwidth]{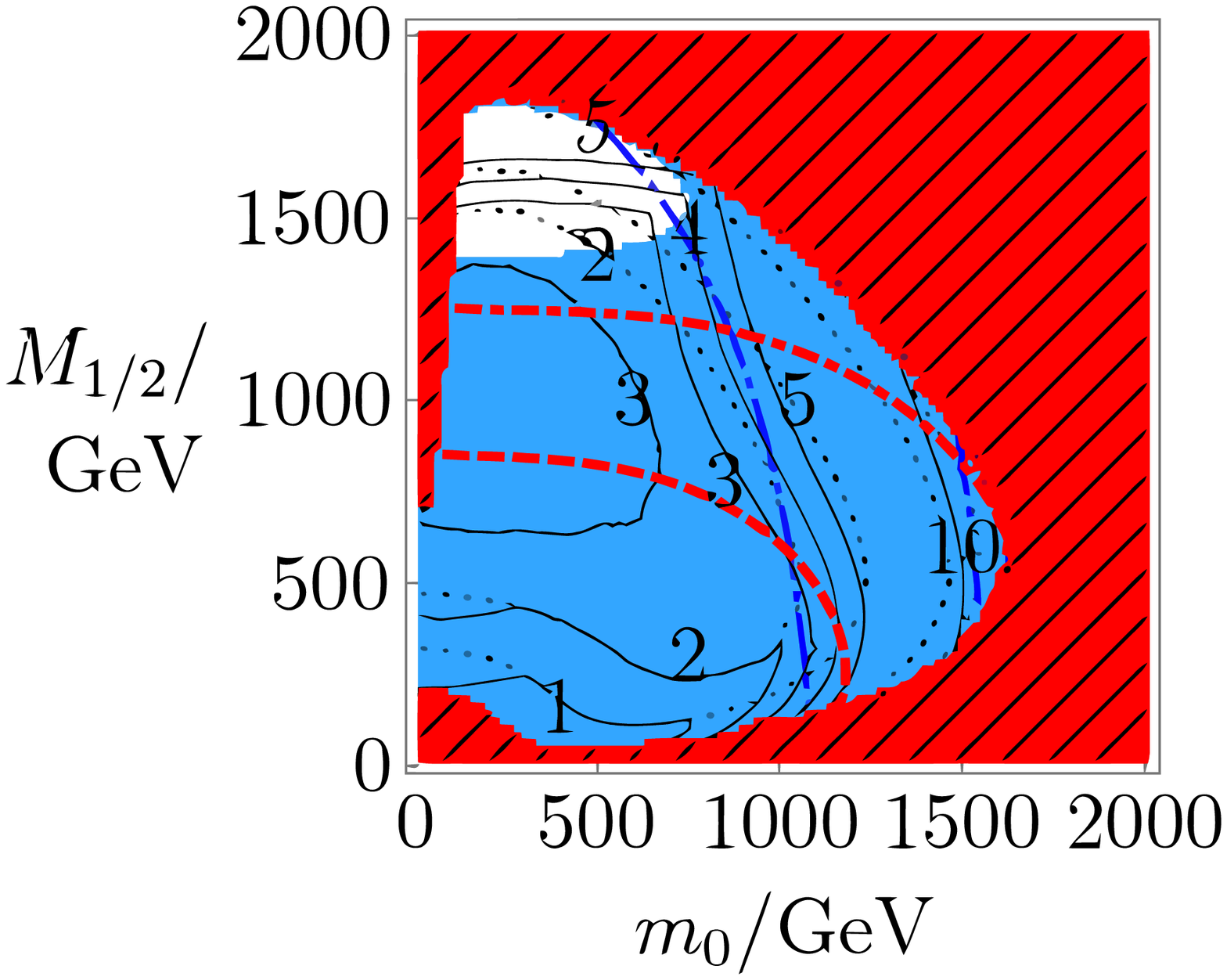}
          }
     \subfigure[770]{
           \label{f.770mFT}
           \includegraphics[width=0.45\textwidth]{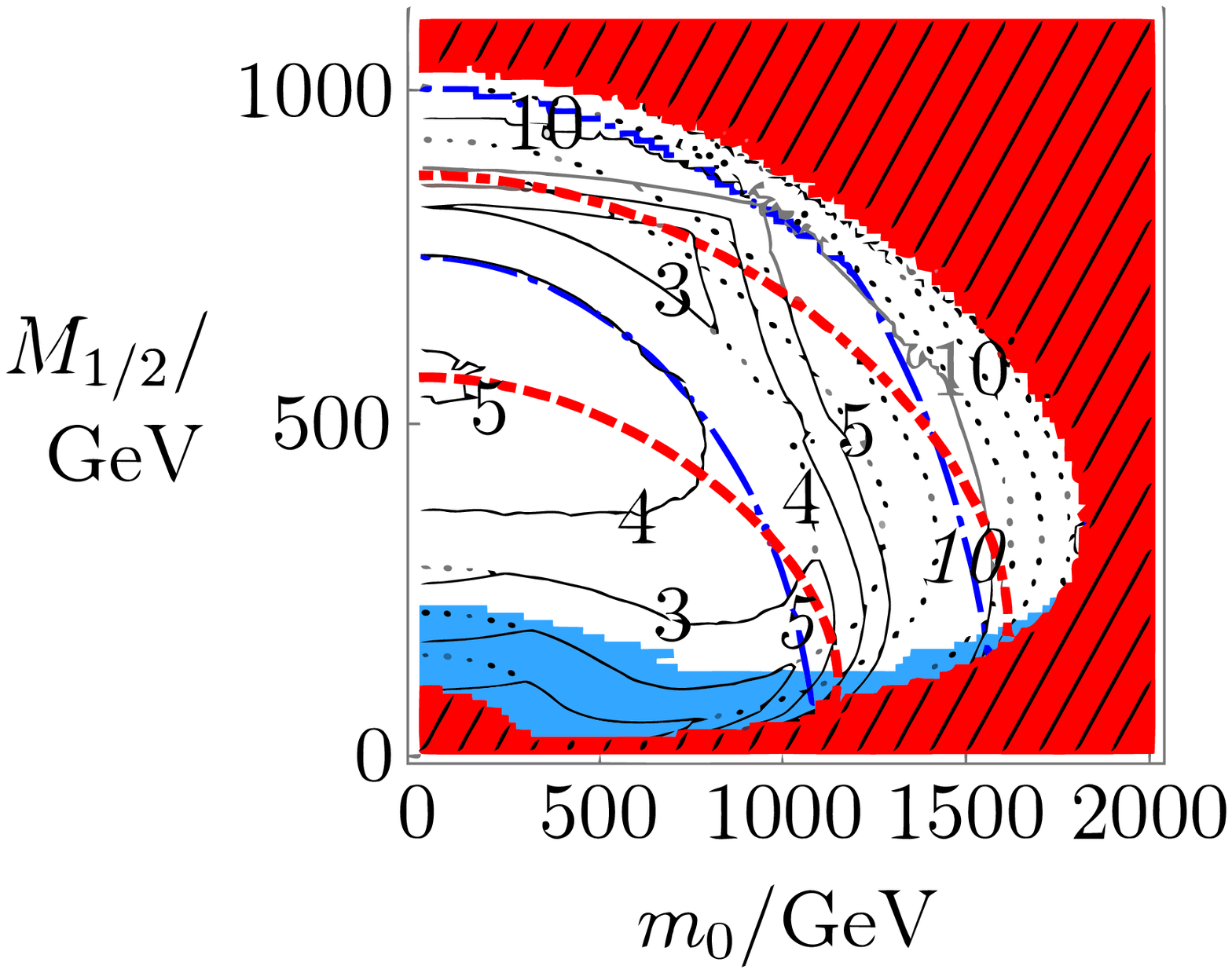}
           }
     \subfigure[O-II]{
           \label{f.OIImFT}
           \includegraphics[width=0.45\textwidth]{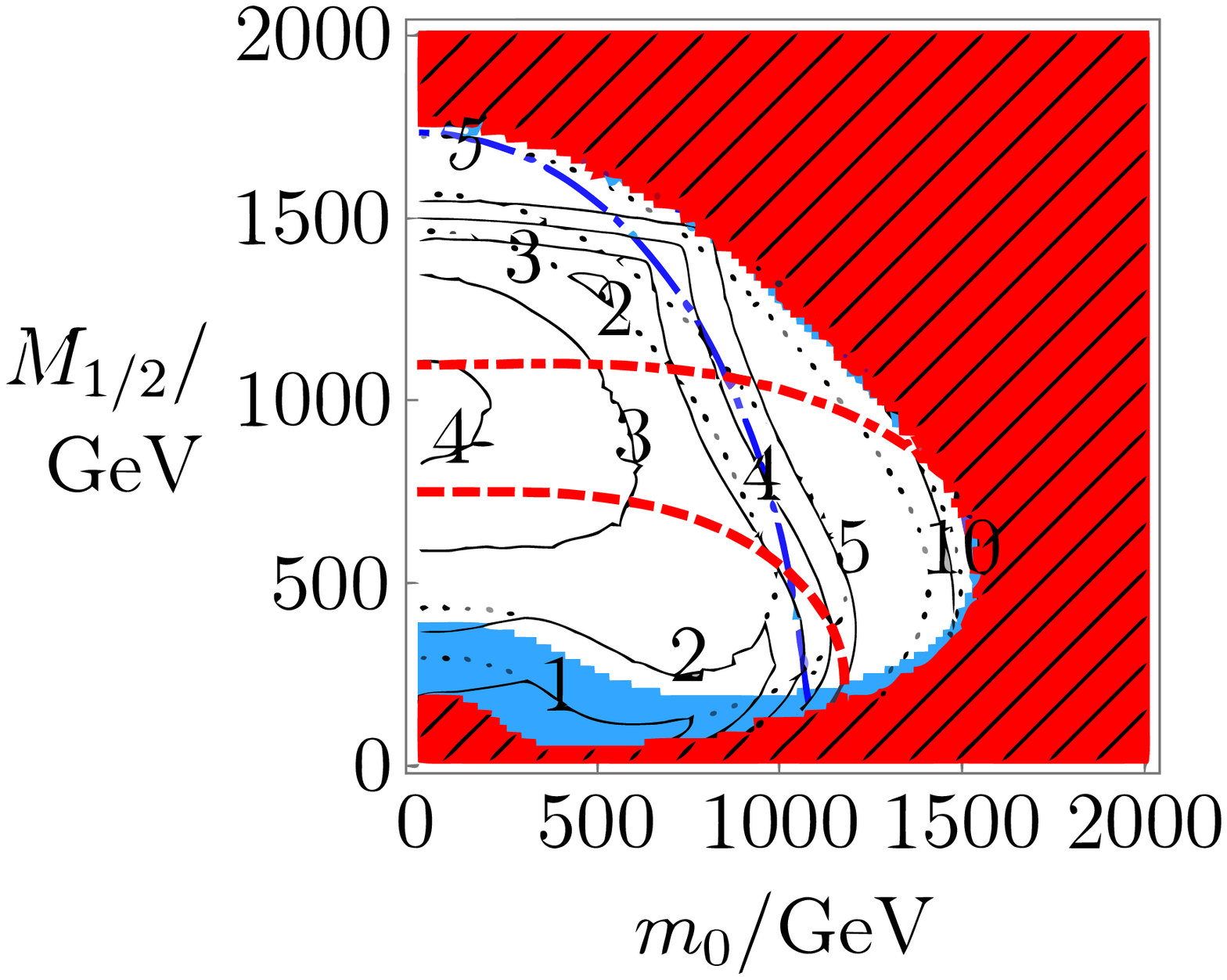}
           }
     \caption{\small These are the same plots as in Fig~\ref{f.FT}, but for the case $\mu<0$. The (blue) dashed contours, with long and short dashing, indicate the sensitivity to the top Yukawa, $\Delta_{h_t}$. The contours for $\Delta_{h_t}$ are labelled in italic.}
     \label{f.FTm}
\end{figure}

We show in Fig~\ref{f.FT} the results of this scan for the O-II, 54, 770 and 210, for $A_0=0$, $\tan\beta=10$ and $\mu>0$. The case $\mu<0$ is shown in Fig~\ref{f.FTm}. Each plot contains several shaded regions: a hatched (red) region where the conditions for successful EWSB are not satisfied; a dark, unhatched (blue) region where the LEP constraints on the masses of superpartners are violated\footnote{In most cases it was the limit on chargino masses that proved to be the most constraining.} and a lighter, unhatched (green) region where the lightest stable particle is charged, which is in conflict with cosmological observations and searches for anomalously heavy molecules \cite{c.CHAMPconstraints}. (We will discuss these last two regions in more detail in Section~\ref{s.Pheno}.) The unbroken lines (black) are contours of constant fine-tuning, $\Delta$, whilst the dotted lines (black) are contours of constant $\Delta_{\mu}$, which we have included so as to `guide the eye', and assist in identifying the regions where $\Delta=|\Delta_{\mu}|$. The dashed and dot-dashed lines (both shown in red) are contours of constant Higgs mass, with $M_{h^0}=111\text{ GeV}$ and $M_{h^0}=114 \text{ GeV}$, respectively.

As can be seen in Figs~\ref{f.FT} and \ref{f.FTm}, EWSB fails to occur in a significant region of the $m_0$---$M_{1/2}$ plane. This is straightforward to understand by considering the structure of the radiative corrections. At the minimum of the one-loop potential the following condition must be satisfied:
\begin{multline}\label{e.MUoneloop}
\lambda^{(0)}v^2+|\mu|^2\approx-z^m m_0^2+\frac{z_{H_d}^mm_0^2+z_{H_d}^MM_{1/2}^2}{\tan^2\beta}
+\frac{3h_t^2}{16\pi^2}\Bigg((m_{Q3}^2+m_{U3}^2+A_t^2)\ln\bigg(\frac{{\rm e}(Q^M)^2}{m_{\tilde{t}_1}m_{\tilde{t}_2}}\bigg)\\
-\bigg[(\mtu^2-\mtl^2)+\frac{(\mtu^2+\mtl^2)}{(\mtu^2-\mtl^2)}A_t^2\bigg]\ln\bigg(\frac{\mtu}{\mtl}\bigg)-2M_t^2\ln\bigg(\frac{m_{\tilde{t}_1}m_{\tilde{t}_2}}{M_t^2}\bigg)\Bigg)
\end{multline}
where here we have only included the dominant $\mathcal{O}(h_t^2)$ contribution. This condition is used to determine $|\mu|$ at each point in parameter space. Note that we have chosen our renormalization scale $Q=Q^M$ such that $z^M=0$, so that the dominant dependence of $\mu$ upon the gaugino mass, $M_{1/2}$, is shifted to the one-loop corrections. For this choice we find that $z^{m}$ is small and positive ($\sim 0.03$), as the scalar focus point scale lies below $Q^M$ in each of these models. It is evident from Eq~\eqref{e.MUoneloop} that as the stop mass scale $M_S^2=m_{\tilde{t}_1}m_{\tilde{t}_2}$ increases and approaches ${\rm e}Q^M$ the contribution of the one-loop corrections to Eq~\eqref{e.MUoneloop} will change sign. This will cause $|\mu|$ to decrease with increasing $M_S$, until eventually $|\mu|<0$. At this point EWSB can no longer occur. The 75 model, which according to Fig~\ref{f.QM} should also exhibit low fine-tuning, is a more extreme example. From Fig~\ref{f.QM} it can be seen that this model has a focus point $Q^M\approx 400 \text{ GeV}$, which is significantly lower than the other models considered here. Because of this, the 75 permits EWSB in only a very small region of the $m_0$---$M_{1/2}$ plane. One can increase the size of this region by permitting non-zero values for $A_0$, which can drive $m_{H_u}^2$ to more negative values. We, however, will not discuss this possibility further here; we simply note that a low scale focus point is not always sufficient to guarantee satisfactory EWSB combined with low fine-tuning.

\begin{figure}[!t]
  \centering
    \subfigure[CMSSM]{
           \label{f.CMSSMFT}
           \includegraphics[width=0.45\textwidth]{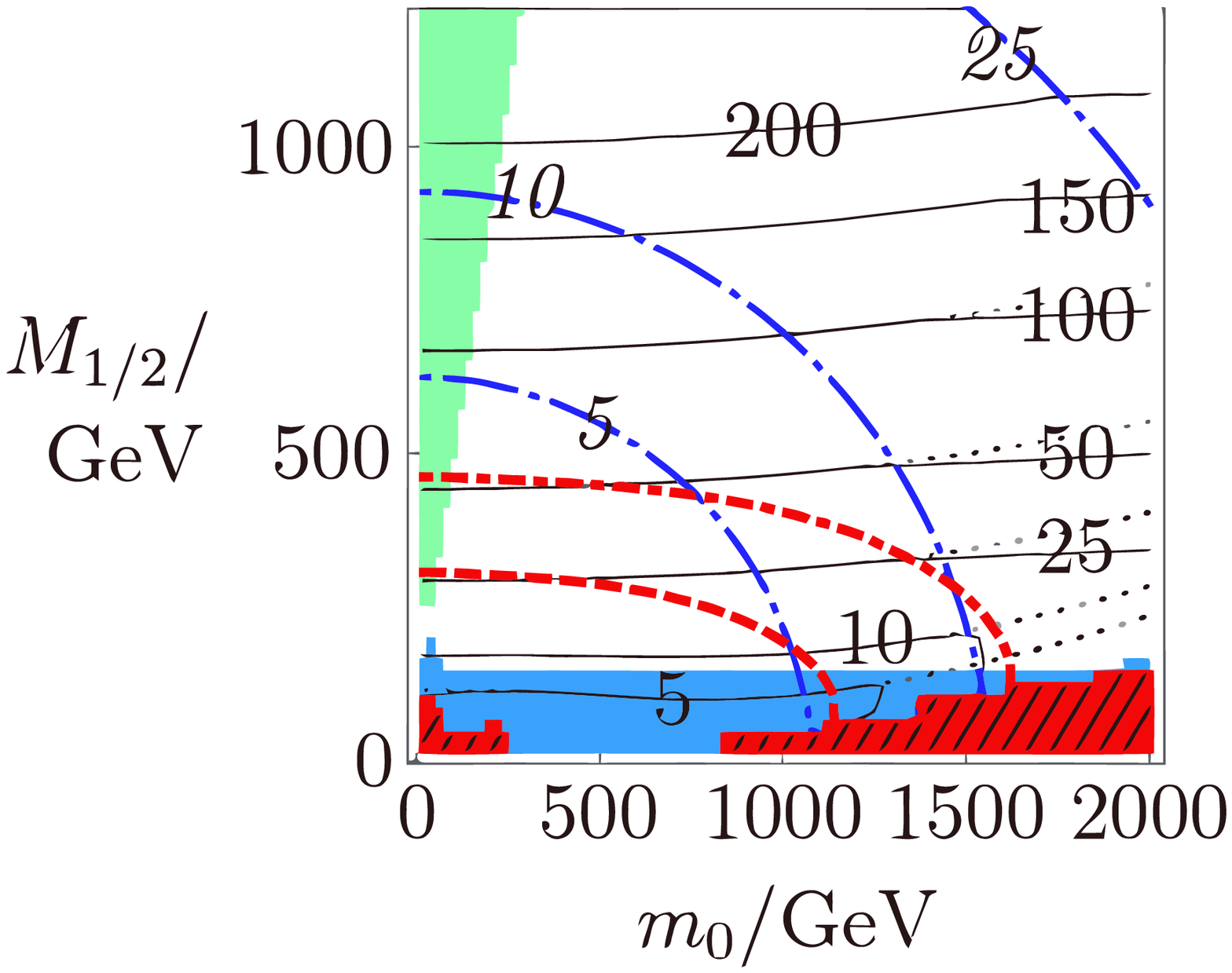}
           }
     \subfigure[$|\Delta_M|$]{
           \label{f.m12FT}
           \includegraphics[width=0.45\textwidth]{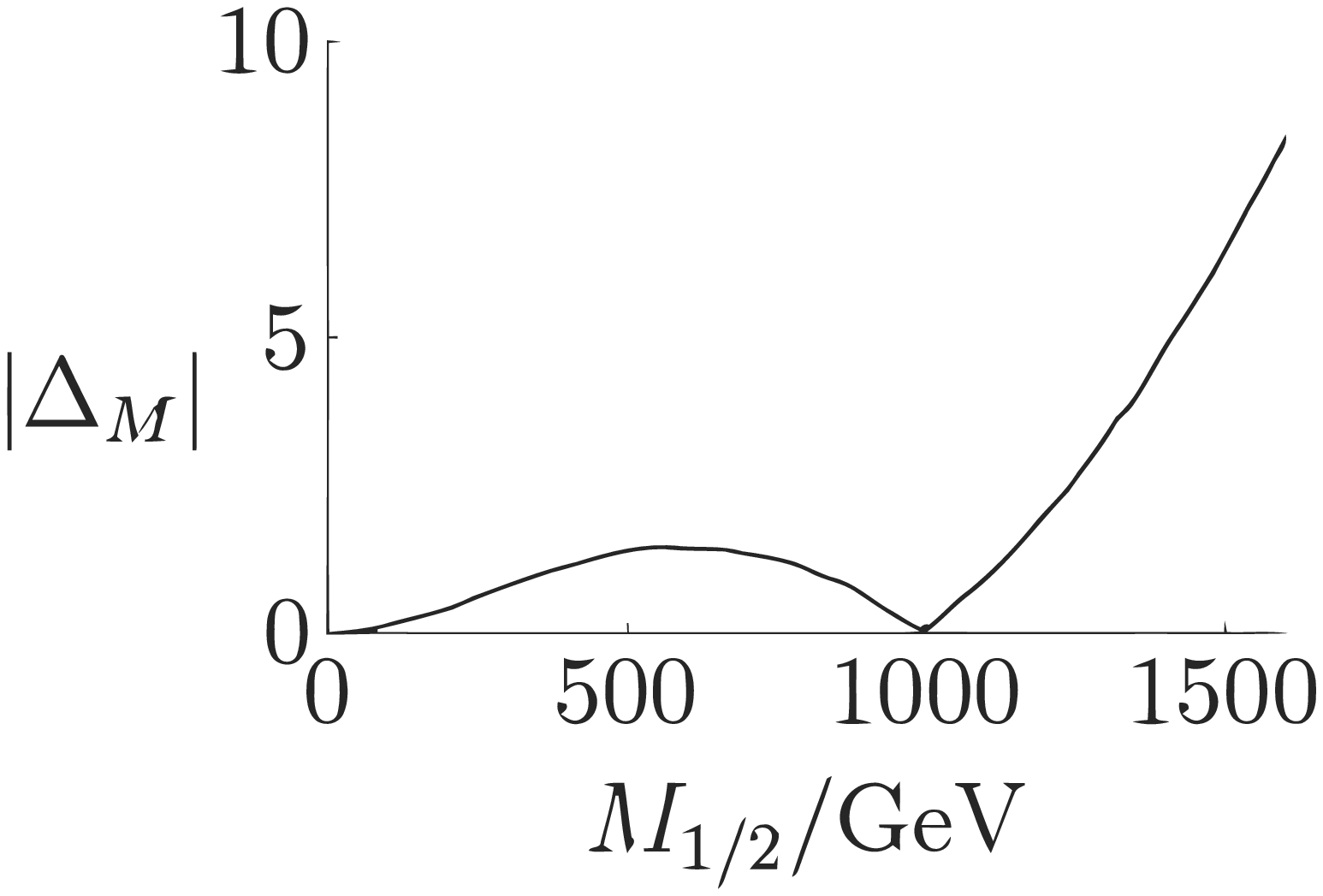}
           }
     \caption{\small Shown in Fig~\ref{f.CMSSMFT} is a plot of the fine-tuning contours in the CMSSM, for $\tan\beta=10$, $A_0=0$ and $\mu>0$. The shaded regions and contours are the same as in Fig~\ref{f.FTm}. Shown in Fig~\ref{f.m12FT} is a plot of $|\Delta_M|$ as a function of $M_{1/2}$, for the O-II model with $\tan\beta=10$, $A_0$=0, $m_0=500\text{ GeV}$ and $\mu>0$.}
\end{figure}

Within the region that permits EWSB each model exhibits a fine-tuning, $\Delta$, which is less than 10. Furthermore, there is a substantial region in which $\Delta<5$. This is to be compared with the CMSSM (which assumes a universal gaugino mass, $\eta_{\alpha}=1$), which is shown in Fig~\ref{f.CMSSMFT} for $\tan\beta=10$ and $A_0=0$. In the CMSSM the region with $\Delta<10$ ($\Delta<5$) is mostly (completely) excluded by constraints from chargino searches at LEP. As discussed in Section \ref{s.GauginoFP}, it is the low scale focus points for both $m_0$ and $M_{1/2}$, present in the 54, 210, 770 and O-II, which suppresses the dependence of $M_Z$ upon these parameters and permits $|\Delta_{m_0}|, |\Delta_{M_{1/2}}|\lesssim 10$ in these models even for $m_0, M_{1/2}\sim 1 \text{ TeV}$.  However, the fine-tunings observed in Figs~\ref{f.FT} and \ref{f.FTm} are somewhat smaller than what one would naively expect from a tree level analysis. For example, at tree level the derivative $\partial v^2 / \partial \mu^2 $ is given by:
\beq
\lambda^{(0)}\frac{\partial v^2 }{\partial \mu^2}=-1+\frac{m_{H_d^2}}{\mu^2\tan^2\beta}
\eeq
Hence the fine-tuning in the $\mu$ parameter is given by:
\beq\label{e.FTmu}
\Delta_{\mu}=\frac{2}{M_Z^2}\bigg(-|\mu|^2+\frac{m_{H_d}^2}{\tan^2\beta}+\mathcal{O}(\frac{|\mu|^2}{\tan^2\beta})\bigg)
\eeq
which, neglecting the second term appearing in Eq~\eqref{e.FTmu}, gives an upper bound of $|\mu|\lesssim 200 \text{ GeV}$ ($|\mu|\lesssim 140 \text{ GeV} $) for regions with $\Delta<10$ ($\Delta<5$). Whilst we find that each model satisfies $|\mu|\lesssim 200 \text{ GeV}$ over the entire $m_0$---$M_{1/2}$ plane, we find that $|\mu|>140 \text{ GeV}$ in regions with $\Delta<5$. There are two reasons why this is so. First, for the case of moderate $\tan\beta=10$, the second term\footnote{The presence of this second term is the direct result of assuming that the soft mass that mixes the Higgs fields is of the form $B\mu H_u H_d$.} appearing in Eq~\eqref{e.FTmu} is comparable to the first for a Higgs soft mass $m_{H_d}\sim 1 \text{ TeV}$. Hence a mild cancellation arises, which tends to reduce $|\Delta_{\mu}|$. Second, when the one-loop corrections are taken into account, the derivative $\partial v^2 / \partial \mu^2$ is given by:
\beq
\lambda^{(0)}(1+\epsilon)\frac{\partial v^2 }{\partial \mu^2}=-1+\frac{m_{H_d^2}}{\mu^2\tan^2\beta}
\eeq
where $\epsilon$ is given by:
\beq
\lambda^{(0)}\epsilon=\frac{3h_t^4}{8\pi^2}\Bigg(\ln\bigg(\frac{m_{\tilde{t}_1}m_{\tilde{t}_2}}{M_t^2}\bigg)+\frac{A_t^2}{m_{\tilde{t}_1}^2-m_{\tilde{t}_2}^2}
\ln\bigg(\frac{m_{\tilde{t}_1}}{m_{\tilde{t}_2}}\bigg)%
\bigg(2-\frac{(m_{\tilde{t}_1}^2+m_{\tilde{t}_2}^2)A_t^2}{(m_{\tilde{t}_1}^2-m_{\tilde{t}_2}^2)^2}\bigg)
+\dotsb
\Bigg)
\eeq
and is related to the one-loop contribution to the Higgs quartic coupling. Hence the fine-tuning is reduced by an overall factor of $(1+\epsilon)\approx M_{h^0}^2/M_{Z}^2$. In regions where the Higgs mass $M_{h^0}\sim 114 \text{ GeV}$, this reduction is a factor between 1.5---2. Thus the large stop masses required to lift the Higgs mass above the LEP bound assist in ameliorating the fine-tuning.

The structure of the radiative corrections, combined with a low focus point scale, reduce the fine-tuning for the other soft parameters in another way. Consider the right hand side of Eq~\eqref{e.MUoneloop} as a function $\mathcal{F}$ of the parameters $a_i$ and $v$ (where here $a_i$ includes soft parameters only),
\beq
\lambda^{(0)}v^2+|\mu^2|\approx \mathcal{F}(a_i,\mu, v^2)
\eeq
In the limit that the soft terms vanish and supersymmetry is restored, $\mathcal{F}$ must vanish. However, as we have noted above, $\mathcal{F}$ will change sign for sufficiently large $M_S$, because of the low scale focus points. Thus $\mathcal{F}$ has another zero for a non-zero value of the soft masses. Between these two zeros the function $\mathcal{F}$ will possess a maximum, at which point the derivative $\partial v^2 / \partial a_i$ will vanish:
\beq
\bigg(\lambda^{(0)}-\frac{\partial\mathcal{F}}{\partial v^2}\bigg)\frac{\partial v^2}{\partial a_i}=\frac{\partial\mathcal{F}}{\partial a_i}=0
\eeq
and hence the fine-tuning will vanish also. We demonstrate this explicitly in Fig~\ref{f.m12FT}, which shows a plot of $|\Delta_{M}|$ in the O-II model, for a scan over $M_{1/2}$ with fixed $m_0=500 \text{ GeV}$. As can be seen, for $M_{1/2}\approx 1 \text{ TeV}$ the fine-tuning vanishes. The tendency of the radiative corrections to reduce the level of fine-tuning is not a new observation, as it has been noted in several previous works. However, in the scenarios considered here, with low scale focus points in both the gaugino and scalar soft masses, the one-loop and tree level contributions to the fine-tuning are comparable, causing this effect to become pronounced.

Let us now discuss the effect of the LEP bound on the Higgs mass. The theoretical calculation of the Higgs mass currently has an uncertainty at the level of $3 \text{ GeV}$\cite{c.HiggsUncertainty}. As a result, we will only consider points in parameter space with $M_{h^0}<111 \text{ GeV}$ to be in violation of the LEP limit. It can be seen in Figs~\ref{f.FT} and ~\ref{f.FTm} that all of the models contain regions of the $m_0$---$M_{1/2}$ plane that can satisfy the LEP constraint, with fine-tunings $\Delta<5$. It is also clear that the LEP bound is far more constraining in the models with $M_3<0$, that is the 54 and 770. This can be understood if one considers the size of the threshold correction $\delta_t$ that, as shown in Eq~\eqref{e.Higgs1loop}, appears in the correction to the Higgs mass. This threshold correction depends upon the degree of stop mixing and is determined by the off diagonal entry in the stop matrix $X_t=A_t-\mu\cot\beta$. In the limit that the stop soft masses are equal, $m_{Q3}^2=m_{U3}^2$,  $\delta_t$ has the following simple form:
\beq
\delta_t=\hat{X}_t^2\bigg(1-\frac{\hat{X}_t^2}{12}\bigg)
\eeq
where $\hat{X}_t=X_t/M_S$. (Although the stop soft masses are not equal in the models we consider, this limit serves to neatly illustrate our point.)
Because $\mu\cot\beta$ is typically small in these models ($\lesssim 20 \text{ GeV}$), $X_t\approx A_t$. For the case $A_t(M_X)=0$ that we consider here, the trilinear term at the low scale is generated solely through radiative corrections. At a scale $Q=500 \text{ GeV}$ the trilinear term is given by:
\beq\label{e.AtLowScale}
A_t(Q=500\text{ GeV})=M_{1/2}\big(-1.61\eta_3-0.25-0.03\eta_1\big)
\eeq
Hence in models with $\eta_3<0$ a cancellation occurs between the first and second terms, which tends to suppress the ratio $\hat{X}_t$. As a result the 54 and 770 models have smaller values for $\delta_t$ (we find that $\delta_t\approx 0.2$ and $0.1$ for the 54 and 770 respectively), compared to the 210 and O-II (which both have $\delta_t\approx 1.1$). From Eq~\eqref{e.StopBound} it thus follows that they need heavier stops in order to satisfy the Higgs bound.

Thus we have identified several models that can satisfy the LEP bound and give the correct EWSB scale without significant fine-tuning in the dimensionful parameters. However, one may worry that the models contain other sources of fine-tuning. Here we will briefly discuss two possibilities: tunings in the dimensionless parameters and tunings that may arise at larger values of $\tan\beta$.

\subsection{Tunings in $h_t(M_X)$}
As pointed out by Romanino and Strumia \cite{c.RomaninoStrumia}, it is difficult to make a case for a model that claims to reduce fine-tuning in the soft terms if it requires a large tuning in its dimensionless parameters---in particular the top Yukawa coupling at the high scale $h_t(M_X)$. 
This parameter is unlike the other dimension-1 parameters, since its value is fixed by low energy observables and the spectrum of superpartners. For the SPS point 1a we find $h_t(M_X)=0.494\pm0.18$, where the error combines uncertainties from the measured values of $M_t$ and $\alpha_S(M_Z)$. To reflect this difference we follow previous works by calculating the fine-tuning in $h_t$ using the following measure:
\beq
\Delta_{h_t}=\frac{\delta h_t}{h_t}\frac{\partial \ln M_Z}{\partial \ln h_t}
\eeq
where $\delta h_t$ is the uncertainty in $h_t(M_X)$. In Fig~\ref{f.FTm} we plot contours of $\Delta_{h_t}$, which are shown with (blue) alternating long and short dashes, assuming that $\delta h_t / h_t = 4 \%$. This fractional uncertainty is slightly larger than any found in the SPS points displayed in Table~\ref{t.Couplings}. We note that for each model $\Delta_{h_t}<10$ across the $m_0$---$M_{1/2}$ plane, which suggests that these models do not require a precise tuning of $h_t$. Furthermore, for models with $\eta_3=1/3$, we note that $\Delta_{h_t}<5$ within most of the region with $\Delta<5$.

\subsection{Fine-tuning at large $\tan\beta$}
For all of the parameter scans presented here we have chosen a fixed value of $\tan\beta=10$. For larger values of $\tan\beta=20$, $30$ and $40$ we have found that the essential structure remains the same; the low scale focus points still permit regions in the $m_0$---$M_{1/2}$ plane that satisfy the Higgs mass bound with low fine-tuning. 
The only significant effect of increasing $\tan\beta$ is the shift of the focus points towards lower energy scales, which is a consequence of the slower running of the soft mass $m_{H_u}$ due to the larger bottom and tau Yukawa couplings. As a result, $m_{H_u}$ is driven to less negative values, which has two implications. First, the magnitude of $|\mu|$ is reduced which, according to Eq.~\eqref{e.FTmu}, also reduces $|\Delta_{\mu}|$. Second, the region of parameter space that permits EWSB is reduced, and vanishes completely for $\tan\beta\approx 50$. Hence the fine-tuning tends to decrease for larger values of $\tan\beta$, however this comes at the cost that EWSB is more difficult to achieve.

\section{Phenomenology}\label{s.Pheno}

One of the main characteristics of a model with a low scale focus point are the ratios, $\eta_1$ and $\eta_3$, of the gaugino masses.
From the running of the gaugino masses it is well known that:
\begin{equation}\label{e.MZGauginoMasses}
\begin{aligned}
\\
\frac{M_i(Q)}{M_{1/2}}&=\eta_i\frac{\alpha_i(Q)}{\alpha_i(M_X)}\\
\\
\end{aligned}
\Rightarrow
\begin{aligned}
\frac{M_1(Q)}{M_2(Q)}&\approx 0.5 \eta_1\\
M_2(Q)&\approx 0.8 M_{1/2}\\
\frac{M_3(Q)}{M_2(Q)}&\approx 2.7 \eta_3\\
\end{aligned}
\end{equation}
where the final equalities hold for the choice $Q\approx1 \text{ TeV}$. Thus one can, quite directly, reconstruct the ratios $\eta_1$ and $\eta_3$ by measuring the mass of these states. By comparing these ratios with the `preferred region' of Fig~\ref{f.QM}, one can test the hypothesis that a focus point is responsible for the little hierarchy between the scales of electroweak and supersymmetry breaking.

Another characteristic, which we identified in the previous section, is the existence of a light Higgsino with $|\mu|\lesssim 200 \text{ GeV}$. In regions of parameter space with $|M_1|, M_2>>|\mu|$, the lightest supersymmetric states will be a chargino and two neutralinos, which will have near degenerate masses. The splitting between the mass of these states is given at tree level by:
\begin{align}
m_{\chi_2^0}-m_{\chi_1^0}&=M_Z^2\bigg(\frac{s_W^2}{M_1}+\frac{c_W^2}{M_2}\bigg)+\mathcal{O}(\frac{M_Z^3}{M_2^2})\\
m_{\chi_1^{\pm}}-m_{\chi_1^0}&=\frac{1}{2}M_Z^2\bigg(\frac{s_W^2}{M_1}+\frac{c_W^2}{M_2}\bigg)+\frac{1}{2}M_Z^2\bigg(\frac{s_W^2}{M_1}-\frac{c_W^2}{M_2}\bigg)\epsilon\sin 2\beta+\mathcal{O}(\frac{M_Z^3}{M_2^2})
\end{align}
where $\epsilon=\mu/|\mu|$. Hence, at tree level, $m_{\chi_1^0}<m_{\chi_1^{\pm}}<m_{\chi_2^0}$. For $M_2,|M_1|>1\text{ TeV}$, this mass splitting $m_{\chi_1^{\pm}}-m_{\chi_1^0}\lesssim 4 \text{ GeV}$. One loop corrections to these masses, which are dominated by third generation squark/quark couplings, generate further contributions that can significantly affect these splittings \cite{c.degenrateHiggsinos}. As can be seen by the (green) light-shaded region of Fig~\ref{f.PHENO}, for $\mu>0$ there are regions of parameter space in the 54, O-II and 770 where, for sufficiently large $M_{1/2}$, the chargino is driven lighter than the neutralino as a result of these radiative corrections. For $\mu<0$ the chargino is instead driven to a higher mass, and becomes heavier than the $\chi_2^0$. The remainder of the SUSY spectrum, comprising the remaining gauginos and scalars, can be significantly heavier than these states, with masses $\gtrsim 1 \text{ TeV}$.

If $|M_1|\lesssim|\mu|$, then an additional light neutralino will also be present in the spectrum. This occurs in the 54 and 210, for which $|\eta_1|<1$. In these models there are regions of parameter space where the LSP is either dominantly Bino or Higgsino, or a mixture of the two. This is significant for the dark matter abundance, as we will discuss in more detail below. For the 210, with $\eta_1=-1/15$ the Bino is very light; indeed one can estimate from Eq~\eqref{e.MZGauginoMasses} that $M_1< 45 \text{ GeV}$ for $M_{1/2}\lesssim 1.5 \text{ TeV}$. When applying the LEP limits on direct particle searches we placed the following requirement on the LSP:
\beq\label{e.LSPlimit}
m_{\chi_1^0}>46 \text{ GeV}
\eeq
in accordance with the PDG \cite{c.PDG}. This limit is violated over much of the 210 parameter space, which is indicated by the dark shading (blue) in Figs~\ref{f.210FT} and \ref{f.210mFT}. However, as pointed out in \cite{c.LightNeutralinos}, the limit of Eq~\eqref{e.LSPlimit} is inappropriate for a neutralino that is light and dominantly bino, since it has suppressed couplings to the gauge bosons. Therefore it is possible that a significantly larger region of the $m_0$---$M_{1/2}$ is consistent with present observations than is indicated. This, however, requires a more detailed phenomenological analysis that is beyond the scope of this work.

The presence of light Higgsino-like charginos and neutralinos has important consequences for a number of observables, in particular for $b\rightarrow s \gamma$, $a_{\mu}$ and the dark matter abundance. We will now consider each of these in more detail.

\subsection{$b\rightarrow s \gamma$}
FCNCs arise at one-loop within the SM, hence new physics that lies near to the weak scale can generate contributions at the same order. As a result, precise measurements of these FCNCs can provide a strong constraint on new physics models. It is well known that the branching ratio $\text{Br}(B\rightarrow X_s \gamma)$ is particularly constraining for models such as the MSSM, because of its precise experimental determination and the improving accuracy of theoretical calculations. Combining data collected by several experiments, the current world average for this branching ratio is given by \cite{c.HFAG}:
\begin{equation}\label{e.BSGexp}
\text{Br}(\overline{B}\rightarrow X_s\gamma)_{E_\gamma>1.6 \text{ GeV}} = (352\pm23\pm9)\times 10^{-6}
\end{equation}
for a photon energy $E_{\gamma}>1.6 \text{ GeV}$ in the rest frame of the $\overline{B}$ meson. The two errors are, respectively, a combined systematic and statistical error, and a systematic error resulting from the lineshape function\footnote{See \cite{c.HFAG} and references therein for further details.}. The theoretical calculation performed within the SM has been completed at NLO, and some progress has been made towards including contributions at NNLO. Currently the best theoretical estimate of this branching ratio at $\mathcal{O}(\alpha_S^2)$ in the SM is \cite{c.bsgSM}:
\beq\label{e.BSGsm}
{\rm Br}(\overline{B}\rightarrow X_s\gamma)^{(\rm SM)}_{E_\gamma>1.6 \text{ GeV}} = (315\pm23)\times 10^{-6}
\eeq
where non-perturbative effects are the dominant source of uncertainty. Supersymmetric models generate additional contributions to this branching ratio. These, which are currently known up to NLO, are dominantly generated by one-loop diagrams involving a top and charged Higgs or a stop and chargino.

\begin{figure}[!t]
  \centering
     \subfigure[54]{
          \label{f.54PHENO}
          \includegraphics[width=0.45\textwidth]{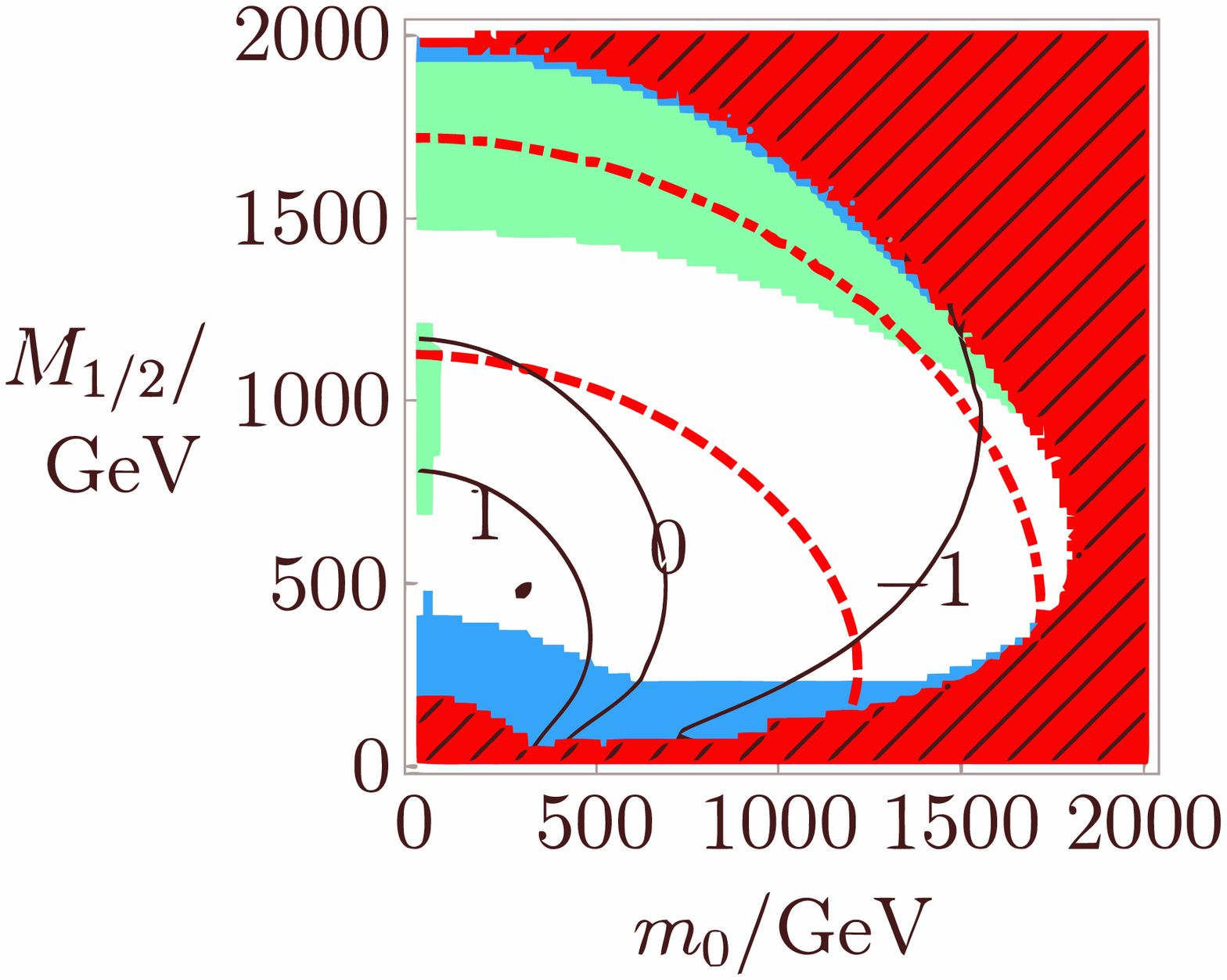}
          }
     \subfigure[210]{
          \label{f.210PHENO}
          \includegraphics[width=0.45\textwidth]{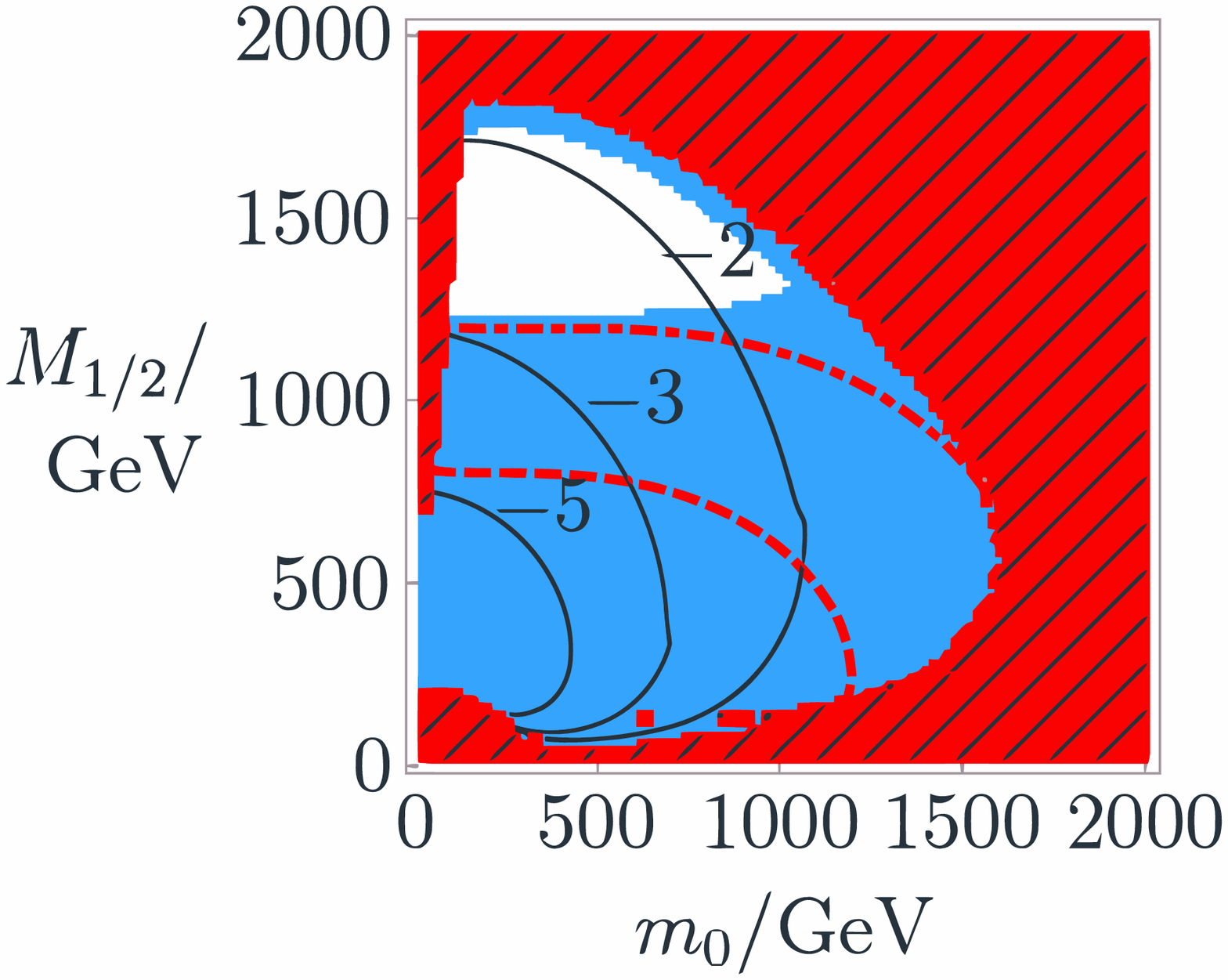}
          }
     \subfigure[770]{
           \label{f.770PHENO}
           \includegraphics[width=0.45\textwidth]{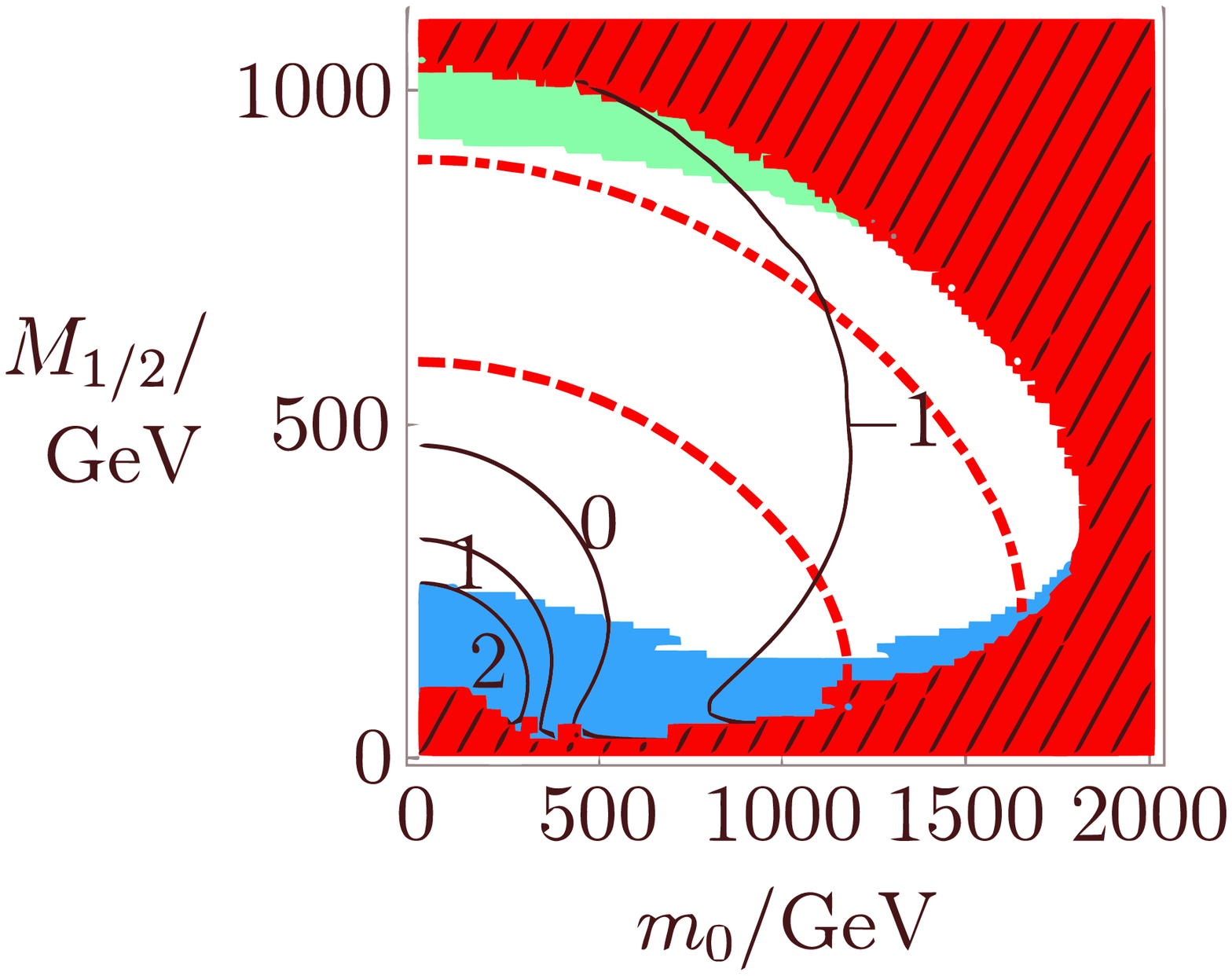}
           }
     \subfigure[O-II]{
           \label{f.OIIPHENO}
           \includegraphics[width=0.45\textwidth]{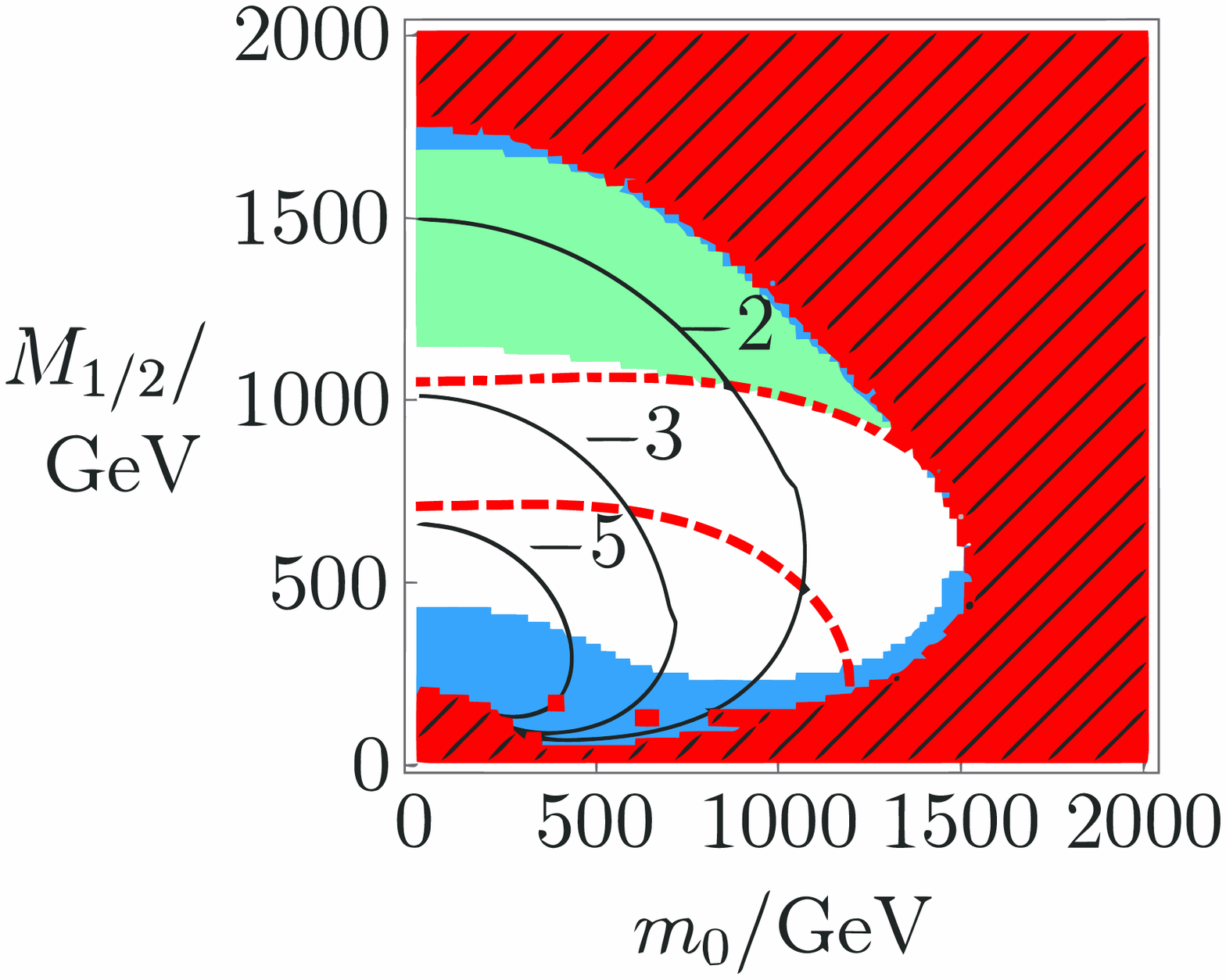}
           }
     \caption{\small Contour plots of the branching ratio ${\rm Br}(\overline{B}\rightarrow X_s \gamma)$ in the $m_0$---$M_{1/2}$ plane. This is for the hypersurface in parameter space with $\tan\beta=10$, $A_0=0$ and $\mu>0$. Solid (black) contours are shown for ${\rm Br}(\overline{B}\rightarrow X_s \gamma)\times 10^4=3.52+ n\sigma $, where $n\in \mathbb{Z}$ and the error $\sigma=0.34$ combines all experimental and theoretical errors in quadrature. The (red) dashed and (red) dot-dashed contours indicate where the Higgs mass, $m_{h^0}$, is 111 GeV and 114 GeV, respectively, whilst the dotted contours correspond to the $|\Delta|=5$ and $|\Delta|=10$ contours from Fig~\ref{f.FT}.}
     \label{f.PHENO}
\end{figure}

\begin{figure}[!t]
  \centering
     \subfigure[54]{
          \label{f.54mPHENO}
          \includegraphics[width=0.45\textwidth]{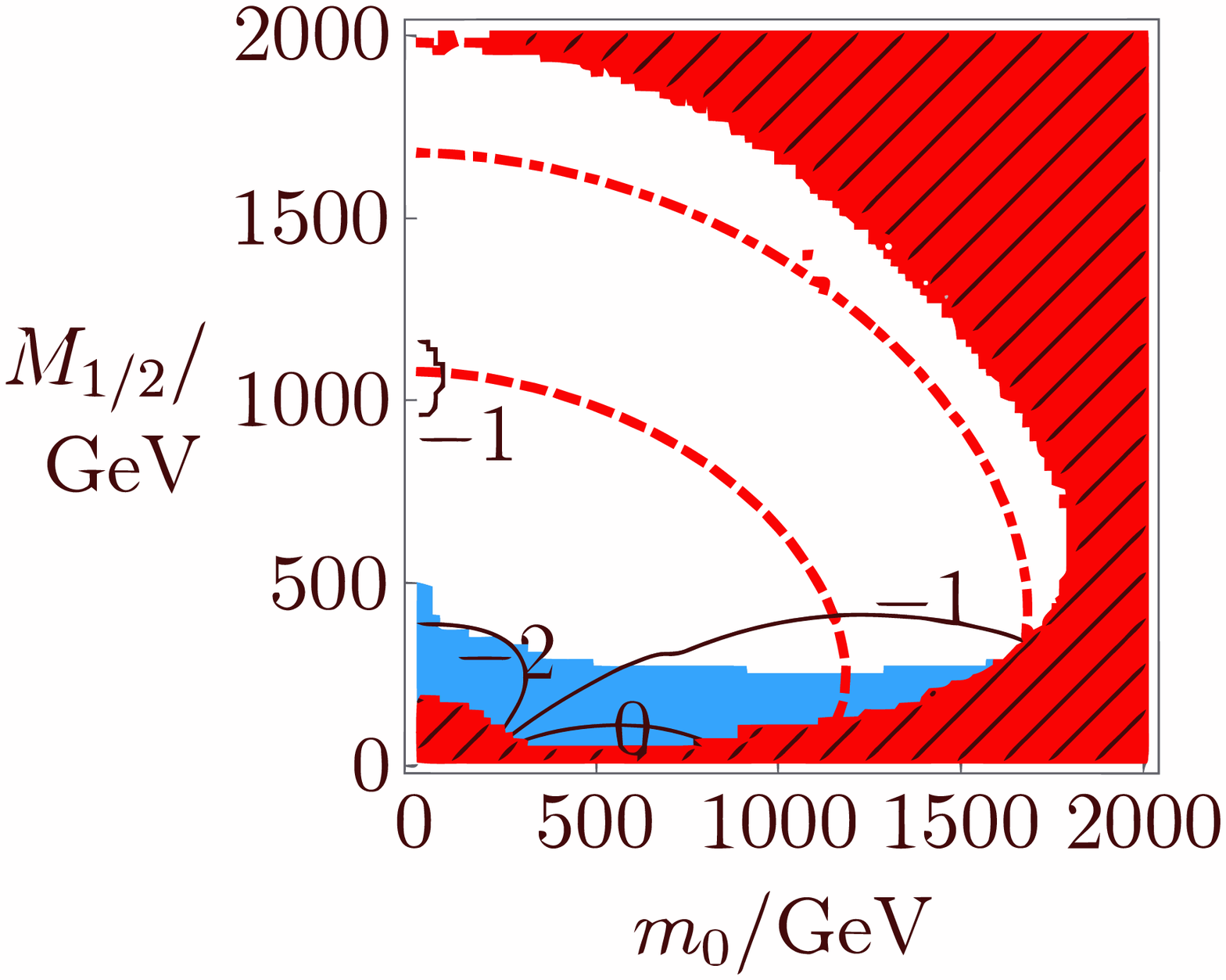}
          }
     \subfigure[210]{
          \label{f.210mPHENO}
          \includegraphics[width=0.45\textwidth]{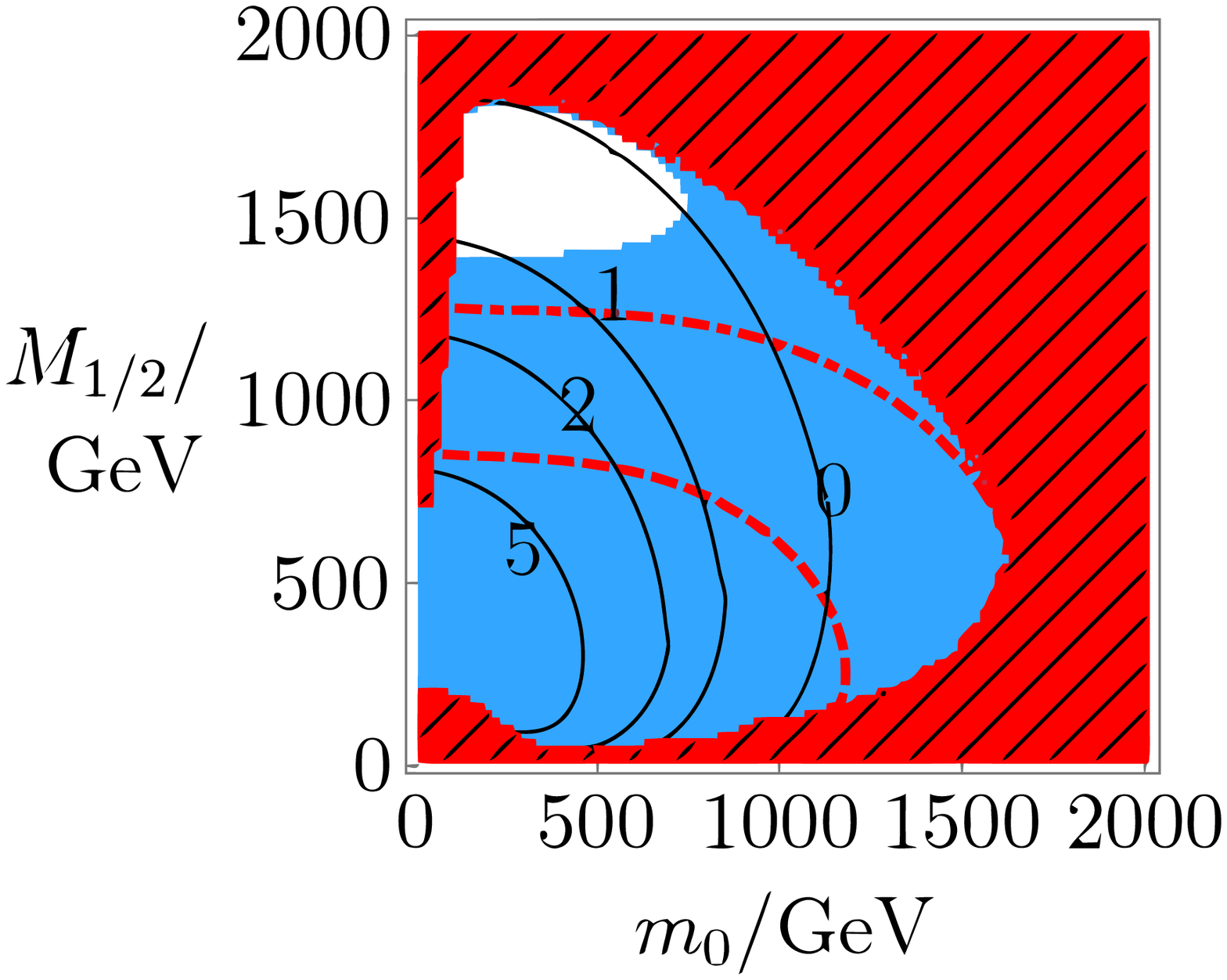}
          }
     \subfigure[770]{
           \label{f.770mPHENO}
           \includegraphics[width=0.45\textwidth]{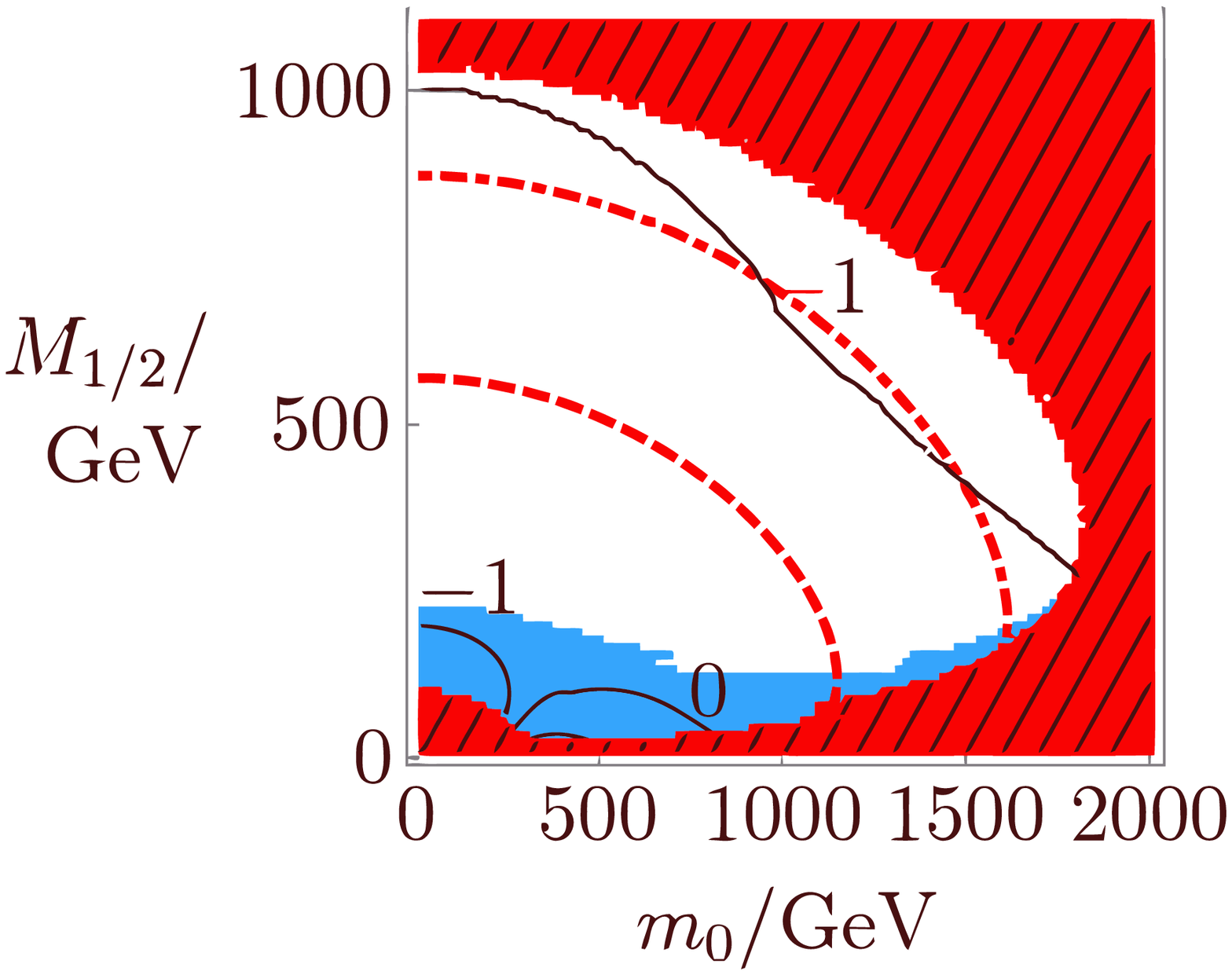}
           }
     \subfigure[O-II]{
           \label{f.OIImPHENO}
           \includegraphics[width=0.45\textwidth]{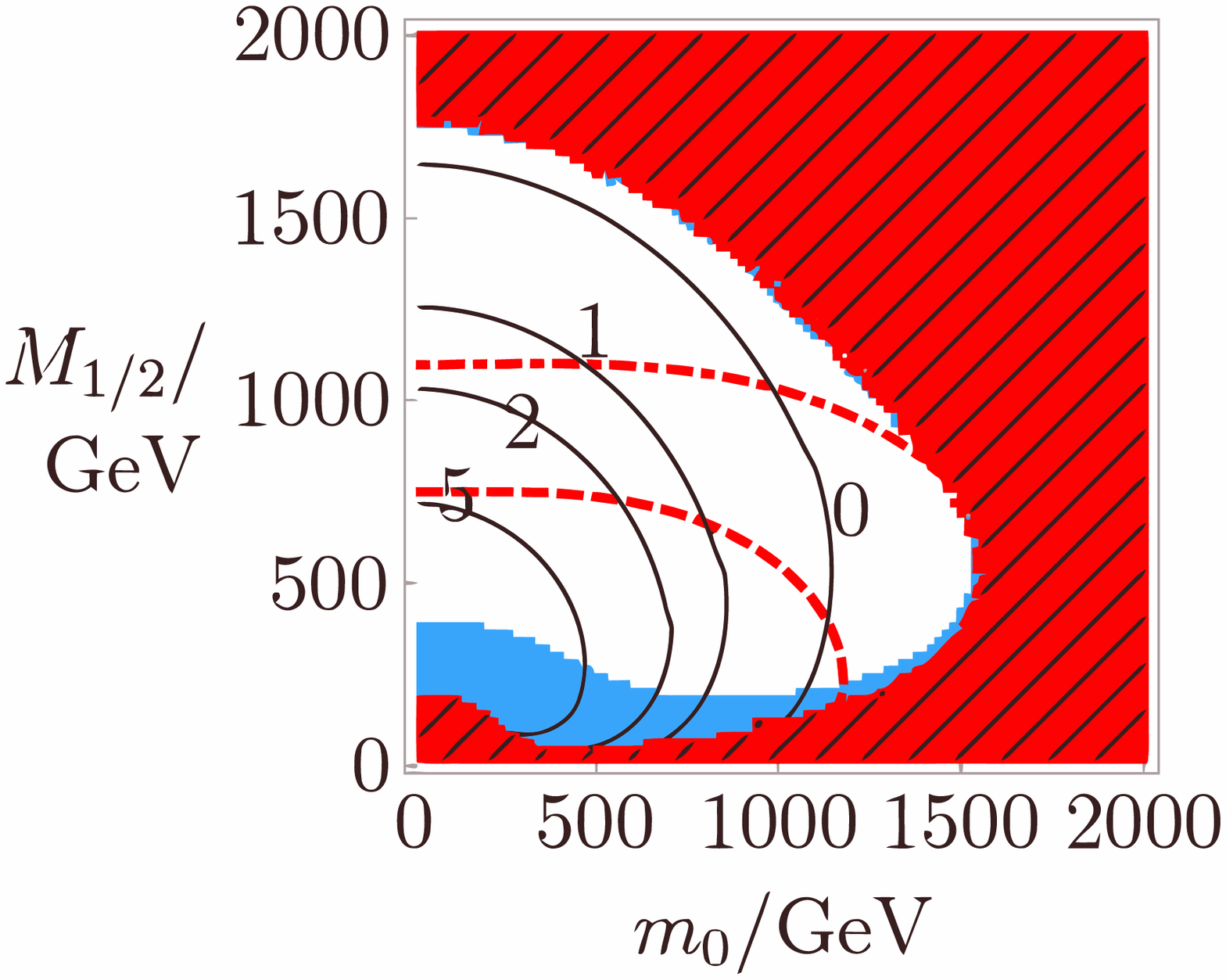}
           }
     \caption{\small These are the same plots as in Fig~\ref{f.PHENO}, but for the case $\mu<0$.}
     \label{f.PHENOm}
\end{figure}

Using the publicly available code {\it superIso} 2.4 \cite{c.superISO}, which includes both the NLO supersymmetric and NNLO contributions from the SM, we have evaluated this branching fraction for each of the models discussed in Section~\ref{s.FT}. The results of this calculation are shown as the solid black contours in Fig~\ref{f.PHENO}, for $A_0=0$, $\tan\beta=10$ and $\mu>0$. The case $\mu<0$ is shown in Fig~\ref{f.PHENOm}. The contours correspond to the $n\sigma$ (for $n\in \mathcal{Z}$) deviation of the prediction from the central experimental value, where $\sigma=34\times 10^{-6}$ combines all experimental and theoretical errors in quadrature. We have used the error in the SM prediction to estimate the theoretical error.

Because the SM prediction is 1$\sigma$ below the experimentally measured value there is a preference for SUSY models that generate a small positive contribution to ${\rm Br}(\overline{B}\rightarrow X_s\gamma)$. The contributions from $H^{\pm}t$ loops are always positive, whilst the sign of loops involving $\chi^{\pm}\tilde{t}$ is given by $\mu A_t$ \cite{c.SignBSG}, which from Eq~\eqref{e.AtLowScale} is given by $- \mu \eta_3$.

The $\chi^{\pm}\tilde{t}$ contribution dominates in the O-II and 210 models, as a result of the small, positive value of $\eta_3$ which, through RGE running, tends to reduce the mass of the stops (relative to the $H^{\pm}$) and give rise to significant stop mixing. As a result there is a tension with the experimentally measured value of ${\rm Br}(\overline{B}\rightarrow X_s\gamma)$ in the $\mu>0$ case, since the $\chi^{\pm}\tilde{t}$ loops generate a large negative contribution which pushes the theoretical prediction to more than 2$\sigma$ below the experimentally measured value over most of the $m_0$---$M_{1/2}$ plane. Hence there is a preference in these models for $\mu<0$, which permits the prediction to lie within 1$\sigma$ of experiment for large values of $m_0$ and $M_{1/2}$, which coincides with the region that generates a sufficiently heavy Higgs.

The 54 and 770, however, have larger, negative values of $\eta_3$, and so tend to possess heavier stops and small stop mixing. Because of this the $H^{\pm}t$ and $\chi^{\pm}\tilde{t}$ contributions are comparable in these models. For the $\mu>0$ case both contributions are positive and as a result there is a significant region in the $m_0$---$M_{1/2}$ plane where the prediction lies within 1$\sigma$ of experiment. In the $\mu<0$ case there is some cancellation between the two contributions which leads to a prediction that is not significantly shifted from that of the SM.

\subsection{$a_{\mu}$}
\begin{figure}[!t]
  \centering
     \subfigure[54]{
          \label{f.54g2PHENO}
          \includegraphics[width=0.45\textwidth]{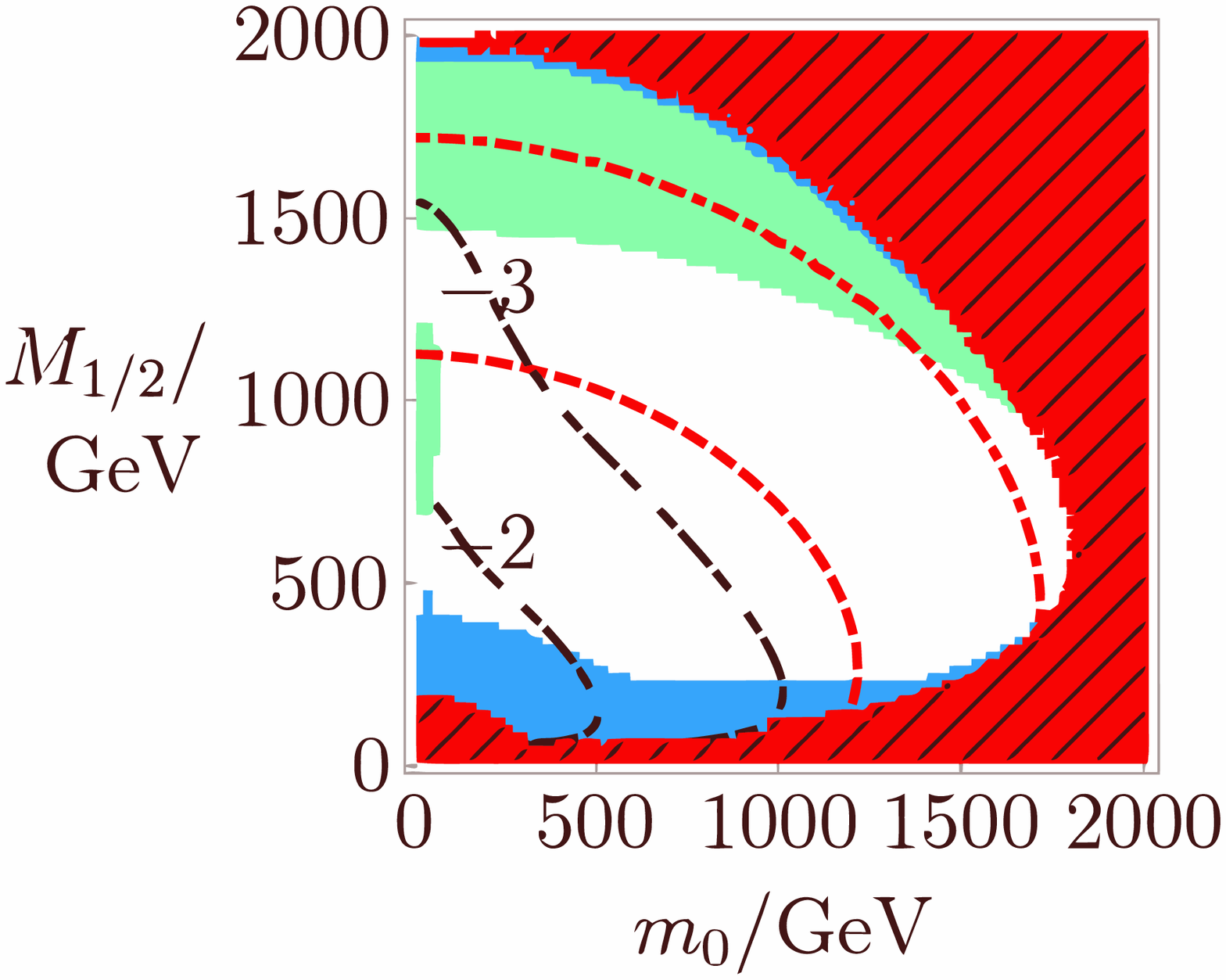}
          }
     \subfigure[210]{
          \label{f.210g2PHENO}
          \includegraphics[width=0.45\textwidth]{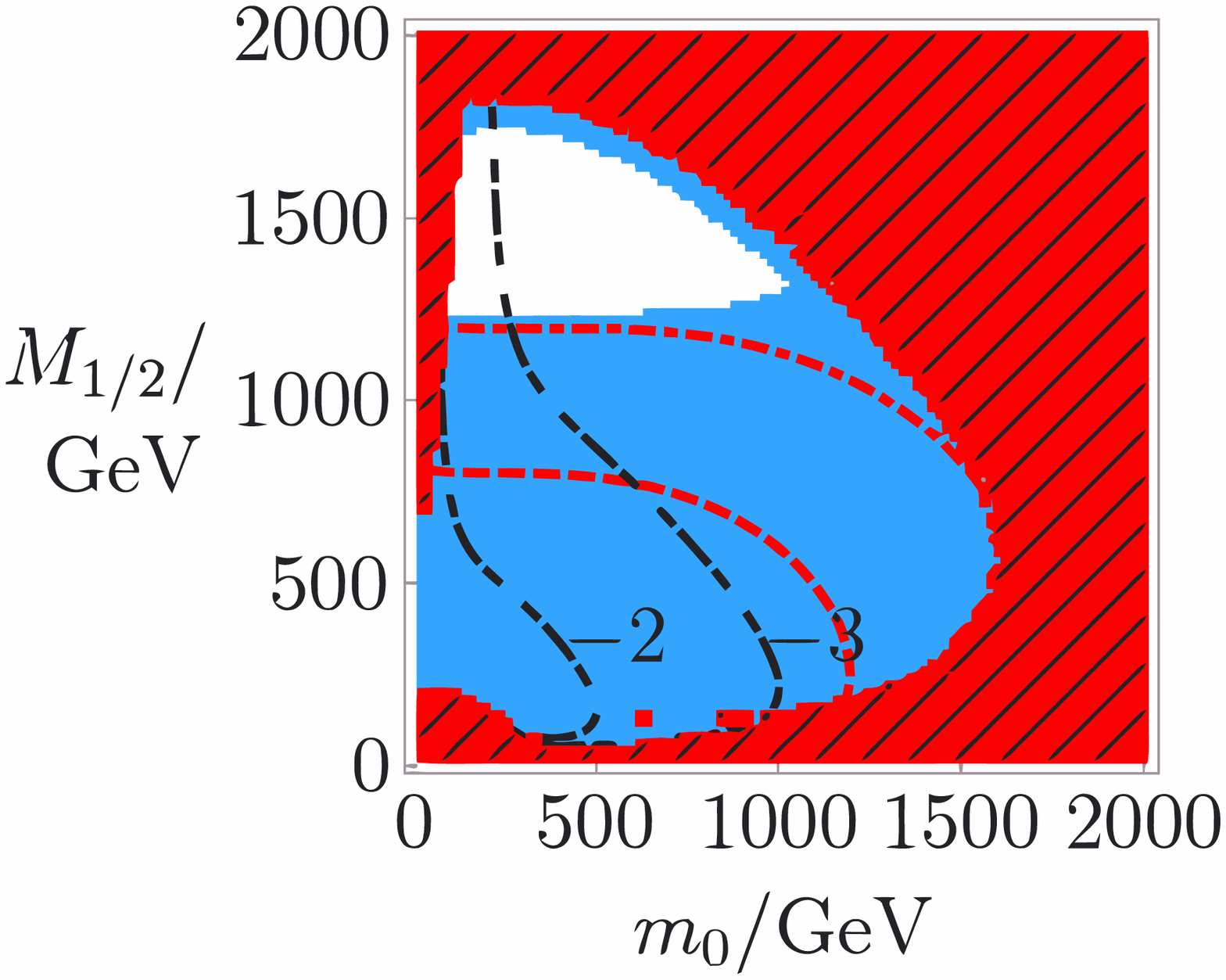}
          }
     \subfigure[770]{
           \label{f.770g2PHENO}
           \includegraphics[width=0.45\textwidth]{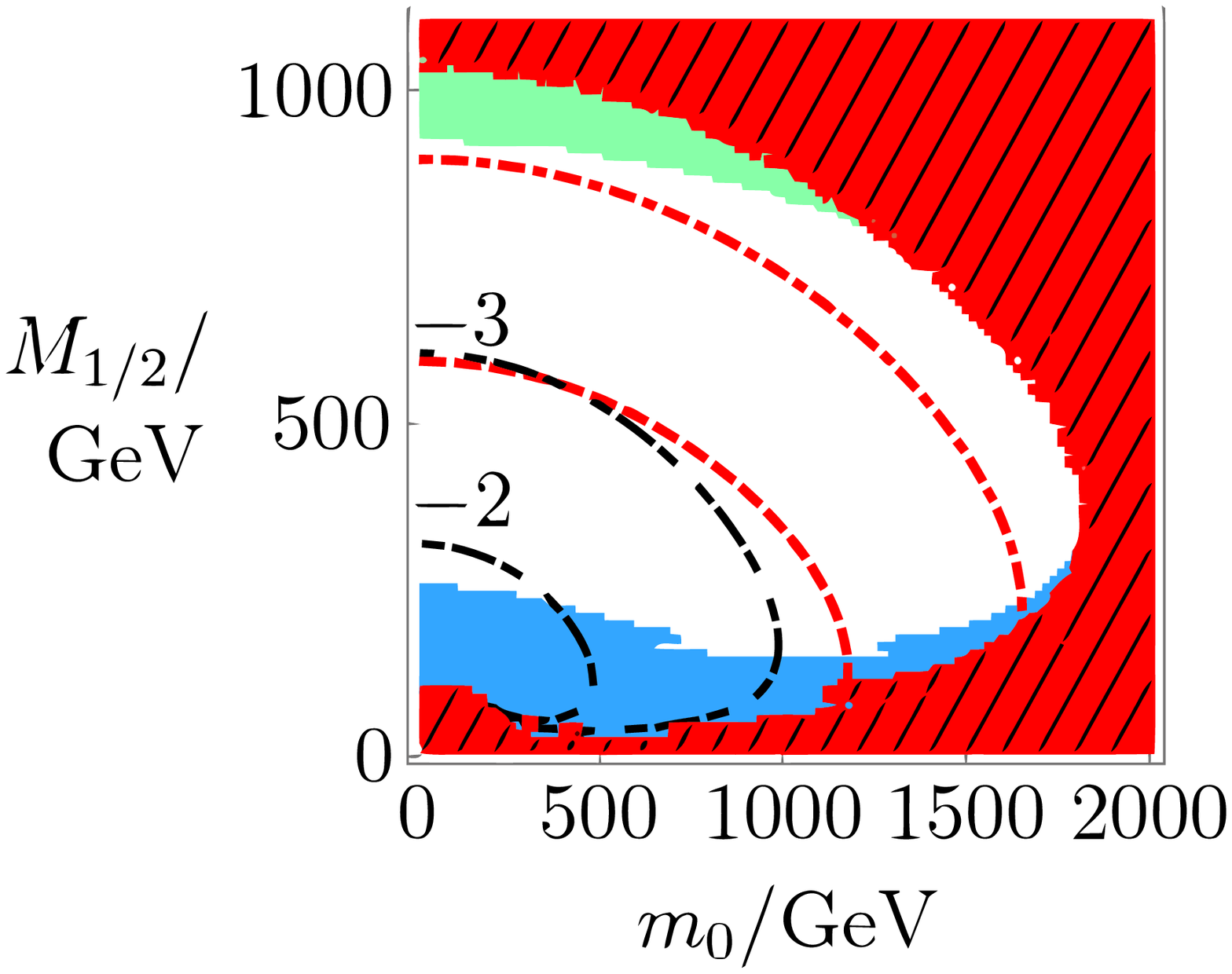}
           }
     \subfigure[O-II]{
           \label{f.OIIg2PHENO}
           \includegraphics[width=0.45\textwidth]{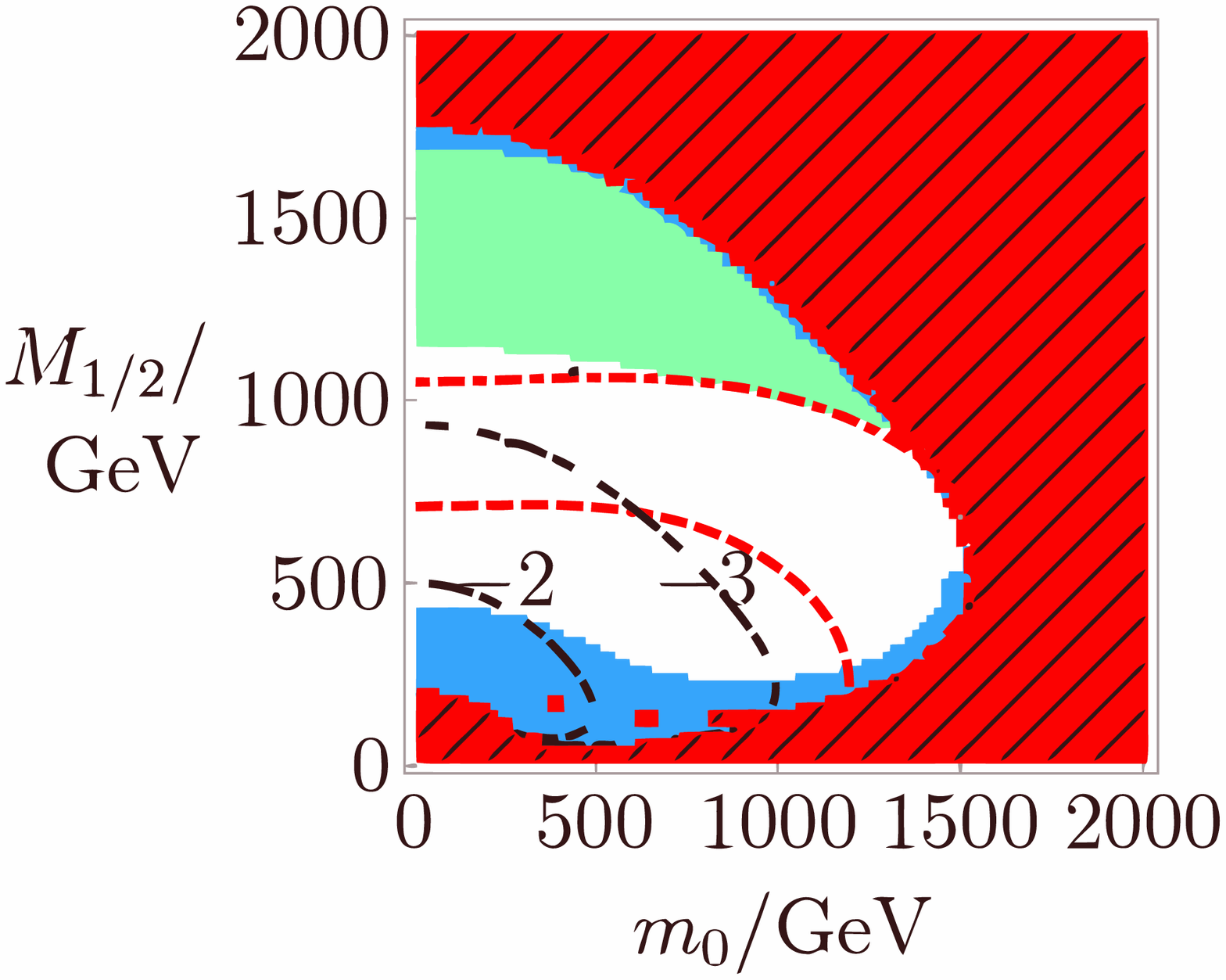}
           }
     \caption{\small Contour plots of $\delta a_{\mu}$, in the $m_0$---$M_{1/2}$ plane. This is for the hypersurface in parameter space with $\tan\beta=10$, $A_0=0$ and $\mu>0$. Dashed (black, with alternating long and short dashing) contours are shown for $\delta a_{\mu}\times 10^{10}=27.5+n\sigma'$, where $n\in \mathbb{Z}$ and the error $\sigma'=8.1$ combines all experimental and theoretical errors in quadrature. The (red) dashed and (red) dot-dashed contours indicate where the Higgs mass, $m_{h^0}$, is 111 GeV and 114 GeV, respectively, whilst the dotted contours correspond to the $|\Delta|=5$ and $|\Delta|=10$ contours from Fig~\ref{f.FT}.}
     \label{f.g2PHENO}
\end{figure}

\begin{figure}[!t]
  \centering
     \subfigure[54]{
          \label{f.54g2mPHENO}
          \includegraphics[width=0.45\textwidth]{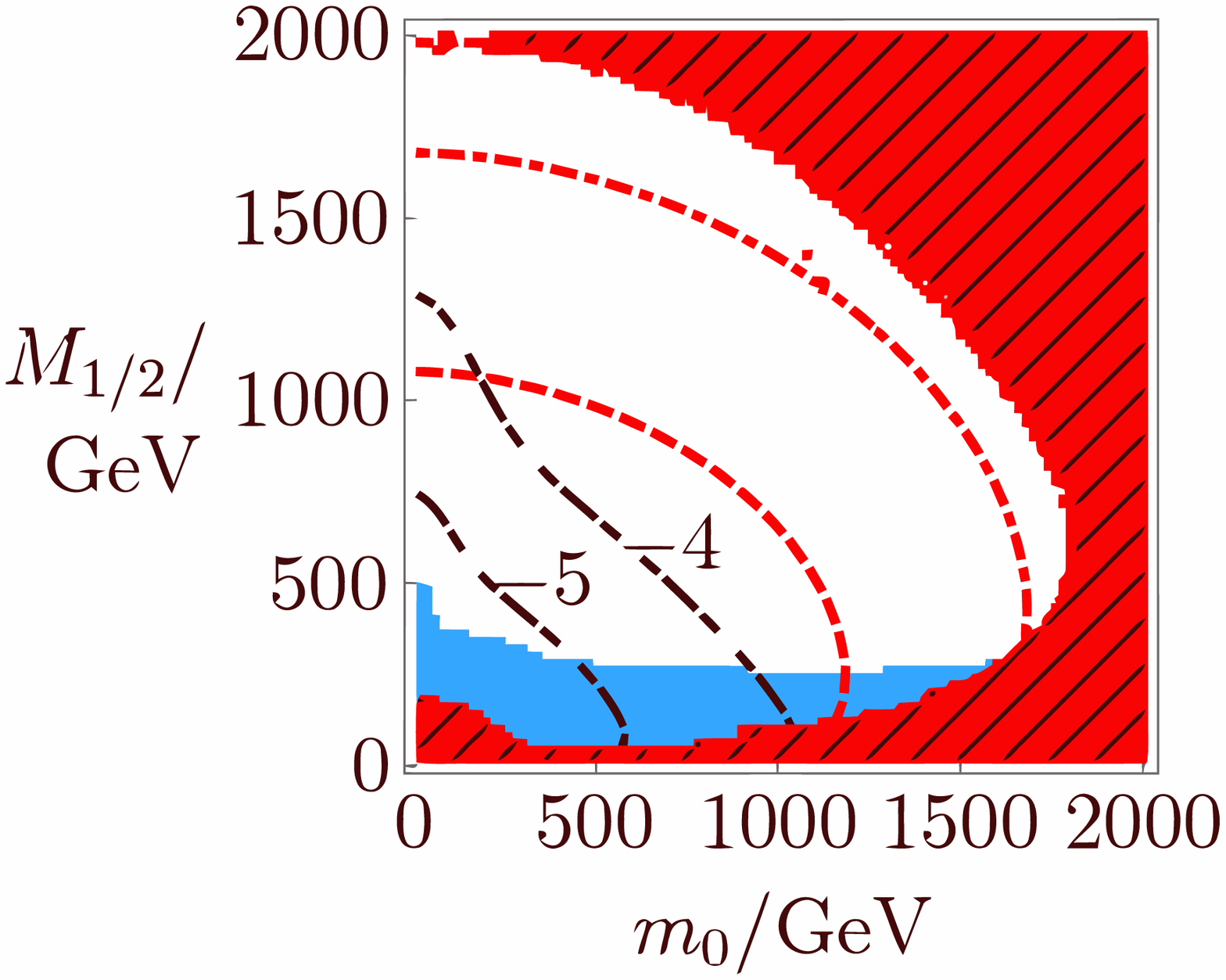}
          }
     \subfigure[210]{
          \label{f.210g2mPHENO}
          \includegraphics[width=0.45\textwidth]{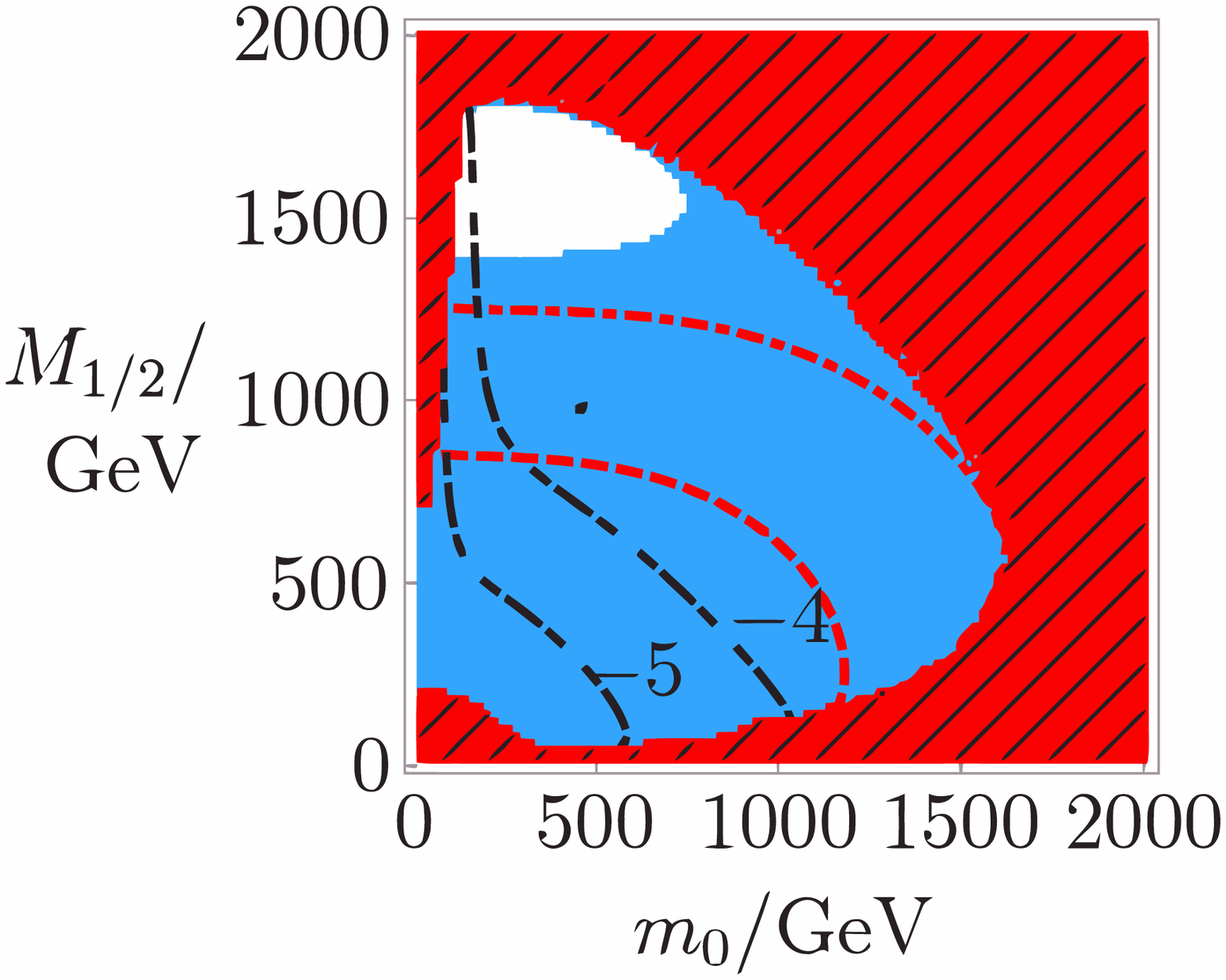}
          }
     \subfigure[770]{
           \label{f.770g2mPHENO}
           \includegraphics[width=0.45\textwidth]{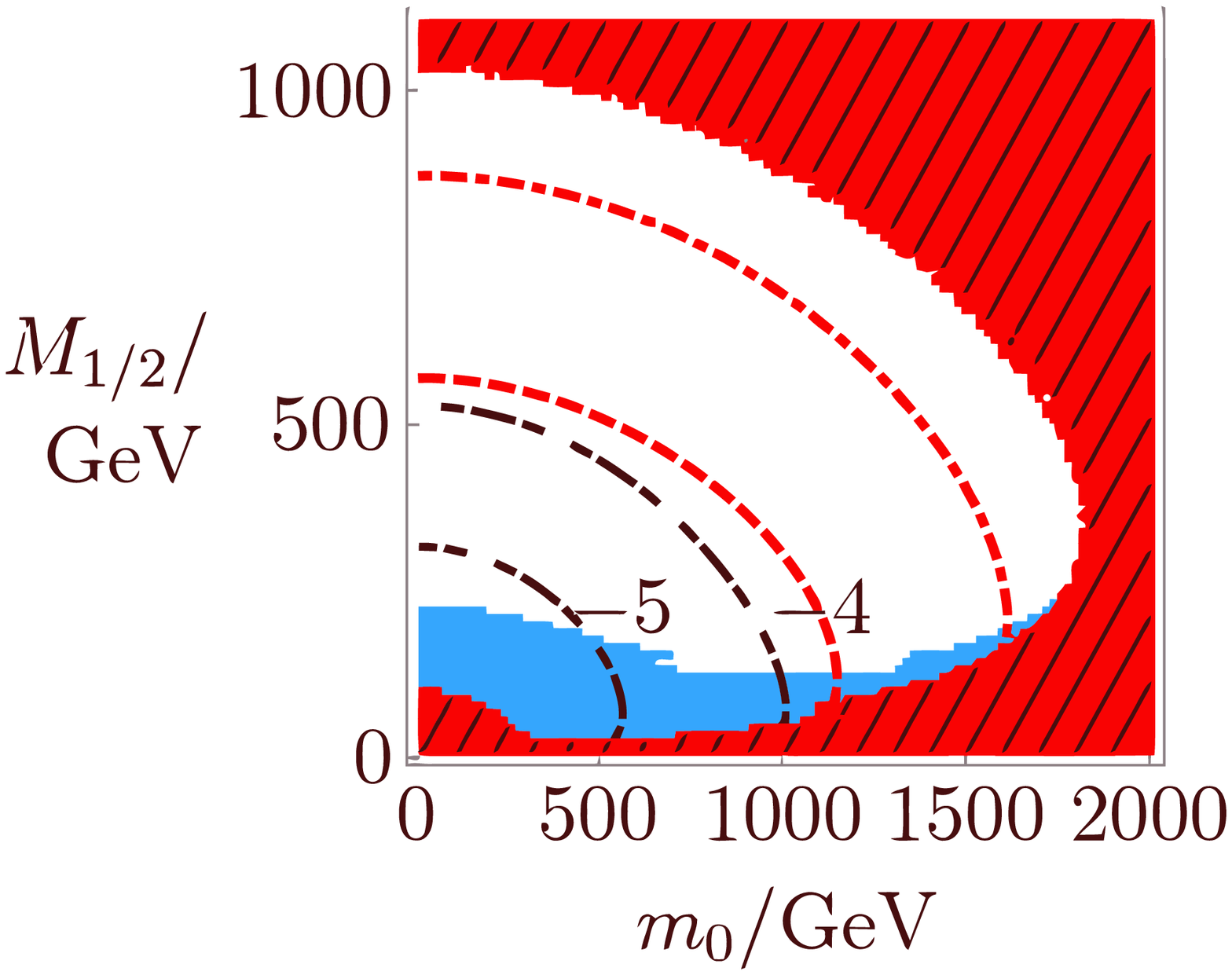}
           }
     \subfigure[O-II]{
           \label{f.OIIg2mPHENO}
           \includegraphics[width=0.45\textwidth]{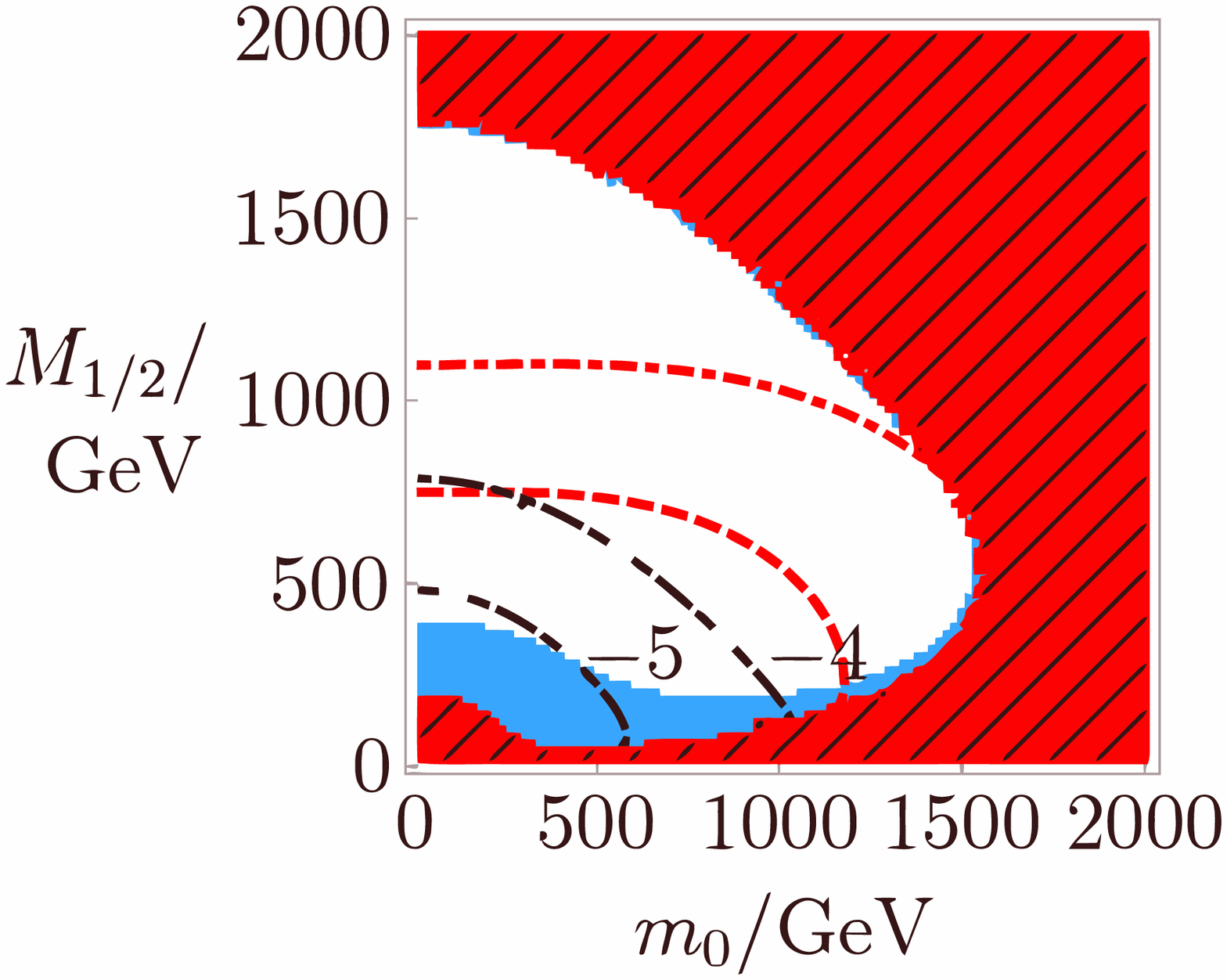}
           }
     \caption{\small These plots are similar to those in Fig~\ref{f.g2PHENO}, except they illustrate the case $\mu<0$.}
     \label{f.g2PHENOm}
\end{figure}

The measurement of the muon's anomalous magnetic moment has been performed to such precision that it is now sensitive to contributions from light supersymmetric states. Hence, potentially, it provides an important constraint upon all SUSY models. The latest experimental result from E821 \cite{c.gMinus2Result}:
\beq\label{e.expAMU}
a_{\mu}^{\text{exp}}=11659208.0(5.4)(3.3)\times 10^{-10}
\eeq
is somewhat larger than the current SM prediction for $a_{\mu}$, calculated using the latest $e^+e^-\rightarrow \pi\pi$ data \cite{c.gMinus2Pred}:
\beq\label{e.SMAMU}
a_{\mu}^{\text{SM}}=11659180.5(5.1)\times 10^{-10}
\eeq
where the error is dominated by uncertainties in the hadronic contributions. The discrepancy between the two results, adding all errors in quadrature, is:
\beq\label{e.aMUdiff}
\delta a_{\mu}\equiv a_{\mu}^{\text{exp}}-a_{\mu}^{\text{SM}}=27.5(8.1)\times 10^{-10}
\eeq
which is a discrepancy of $3.4\sigma$. There are a number of unresolved issues with this discrepancy, particularly regarding the true size of the hadronic contribution\footnote{If the hadronic contribution is calculated using data from $\tau$-decays, instead of $e^+e^-$ data, the discrepancy is substantially reduced to $1.4 \sigma$. See \cite{c.aMUdiscrep} for further discussion.}, which has a substantial impact upon the theoretical prediction. Despite this, it is interesting to consider the contribution to $a_{\mu}$ from supersymmetric states. We have evaluated this using a routine included in {\it superIso}, and the results are shown as the dashed contours in Figs~\ref{f.g2PHENO} and \ref{f.g2PHENOm}.

The dominant contribution to $\delta a_{\mu}$ arises from one-loop diagrams involving charginos, neutralinos and slepton states. One can see in Figs~\ref{f.g2PHENO} and \ref{f.g2PHENOm} that the shape of the contours in the low $m_0$ region differs somewhat in models with $\eta_1<0$ (54 and 210) compared to those with $\eta_1>0$ (770 and O-II). This is the result of chargino and neutralino contributions having signs $\mu M_2$ and $-\mu M_1$ respectively; hence a partial cancellation occurs when $M_1M_2>0$. It is clear that the regions which satisfy the Higgs bound typically generate contributions to $a_{\mu}$ that are too small to account for the discrepancy observed in \eqref{e.aMUdiff}. This is caused by the relative heaviness of slepton states in these models, which is a consequence of the large values of either $M_{1/2}$ or $m_0$ required to satisfy the Higgs bound, combined with our assumption of scalar mass universality. Until the uncertainties in the $\delta a_{\mu}$ result are resolved, one cannot say whether these models are already excluded by this result. If the discrepancy between the SM and experiment persists then one may consider further model building in order to account for this. For example, one may consider introducing mild hierarchies into the soft masses by splitting the first two generations from the third. This would permit spectra with the light sleptons necessary for a significant contribution to $\delta a_{\mu}$. However, without a concrete model that can generate such large inter-generational splittings we will not pursue this further here.

\subsection{$\Omega_{\rm CDM}$}

The current best-fit of the $\Lambda{\rm CDM}$ model to the combined data sets of: the five year WMAP observations, measurements of Type-1a supernovas and baryon acoustic oscillations, has determined the following average for the density of cold dark matter (CDM) \cite{c.WMAP5yr1,c.WMAP5yr2}:
\beq\label{e.OmegaCDM}
\Omega_{\rm CDM}h^2=0.1131\pm0.0034
\eeq
Despite this and other compelling evidence for its existence, the precise nature and composition of dark matter are unknown. However, the lightest supersymmetric particle (LSP), which is stable under the assumed R-parity symmetry, can be generated in the early universe and, provided it is a neutralino, is a viable candidate for one component of the observed cold dark matter abundance. Consistency with observations and the $\Lambda{\rm CDM}$ cosmology therefore requires that the neutralino abundance, $\Omega_{\chi}$, is less than $\Omega_{\rm CDM}$. Under the assumption that neutralinos are only produced through the mechanism of thermal freeze-out, and assuming a standard thermal history of the universe, the abundance $\Omega_{\chi}$ depends only upon the spectrum of supersymmetric states. Hence this upper limit provides an important constraint on the allowed regions of parameter space.

Schematically, $\Omega_{\chi}\propto \langle \sigma v \rangle^{-1}$, where $\langle \sigma v \rangle$ is a thermally averaged cross-section \cite{c.EarlyUni} that combines the cross-sections for annihilation of the $\chi_1^0$ and also its co-annihilation with any other supersymmetric states present at the time of freeze-out. Because the lightest neutralino is a superposition of several gauge eigenstates:
\beq
\chi_1^0=N_1^B\widetilde{B}+N_1^W\widetilde{W}^3+N_1^u\widetilde{H}_u^0+N_1^d\widetilde{H}_d
\eeq
its couplings, and therefore $\Omega_{\chi}$, depend critically upon the neutralino's composition, i.e. the coefficients $N_1^i$ that are determined by the neutralino mass matrix.

As we have already discussed, models with a low scale focus point contain a light Higgsino.
In models with $|\eta_1|>1$, as in the O-II and 770, $M_1, M_2\gg|\mu|$ over most of parameter space, and so the lightest neutralino becomes a (near) pure Higgsino state. Because of its gauge interactions the Higgsino can annihilate directly into gauge bosons, whilst the near degeneracy of the lightest neutralino and chargino gives rise to co-annihilation. An analytic calculation performed in \cite{c.WellTempNeut} gives the abundance of a Higgsino pure state as:
\beq\label{e.OmegaHiggsino}
\Omega_{\chi}h^2=0.1\big(\frac{\mu}{1\text{ TeV}}\big)^2
\eeq
hence $\Omega_{\chi}h^2\lesssim 4\times10^{-3}$ over the entire parameter space of these models, which is safely within the bound given by \eqref{e.OmegaCDM}. Thus without positing additional (non-thermal) mechanisms of production, the Higgsino LSP can only give a subdominant contribution to the total cold dark matter abundance. The prospects for the detection of this Higgsino-like component have been studied previously in the context of non-universal gaugino masses, and we refer the interested reader to the literature \cite{c.HiggsinDM}.

For the alternative case that $|\eta_1|<1$, as in the 54 and 210, there exist regions of parameter space where both Higgsino and Bino are light and hence where the lightest neutralino is a mixture of these two eigenstates. The couplings of the Bino are such that it may only annihilate through t-channel exchange of superpartners, such as sleptons or squarks. Hence the annihilation cross-section of a pure Bino LSP becomes suppressed as the masses of these scalars increases, resulting in a larger abundance. As was shown by Arkani-Hamed et. al. \cite{c.WellTempNeut}, for slepton masses $m_{\tilde{e}_R}\gtrsim 111 \text{ GeV}$ the abundance is already forced above the limit set by Eq~\eqref{e.OmegaCDM}. As is demonstrated in Figs~\ref{f.PHENO} and \ref{f.PHENOm}, the constraints from LEP2 and ${\rm Br}(b\rightarrow s \gamma)$ already excludes most of this 'bulk region' of parameter space that permits these light scalars. Thus an acceptable dark matter abundance requires that the cross-section is enhanced either by a significant Higgsino component of the LSP, or through coannihilation or resonance effects.
\begin{figure}[!t]
  \centering
     \subfigure[54, $\mu>0$]{
          \label{f.54DM}
          \includegraphics[width=0.45\textwidth]{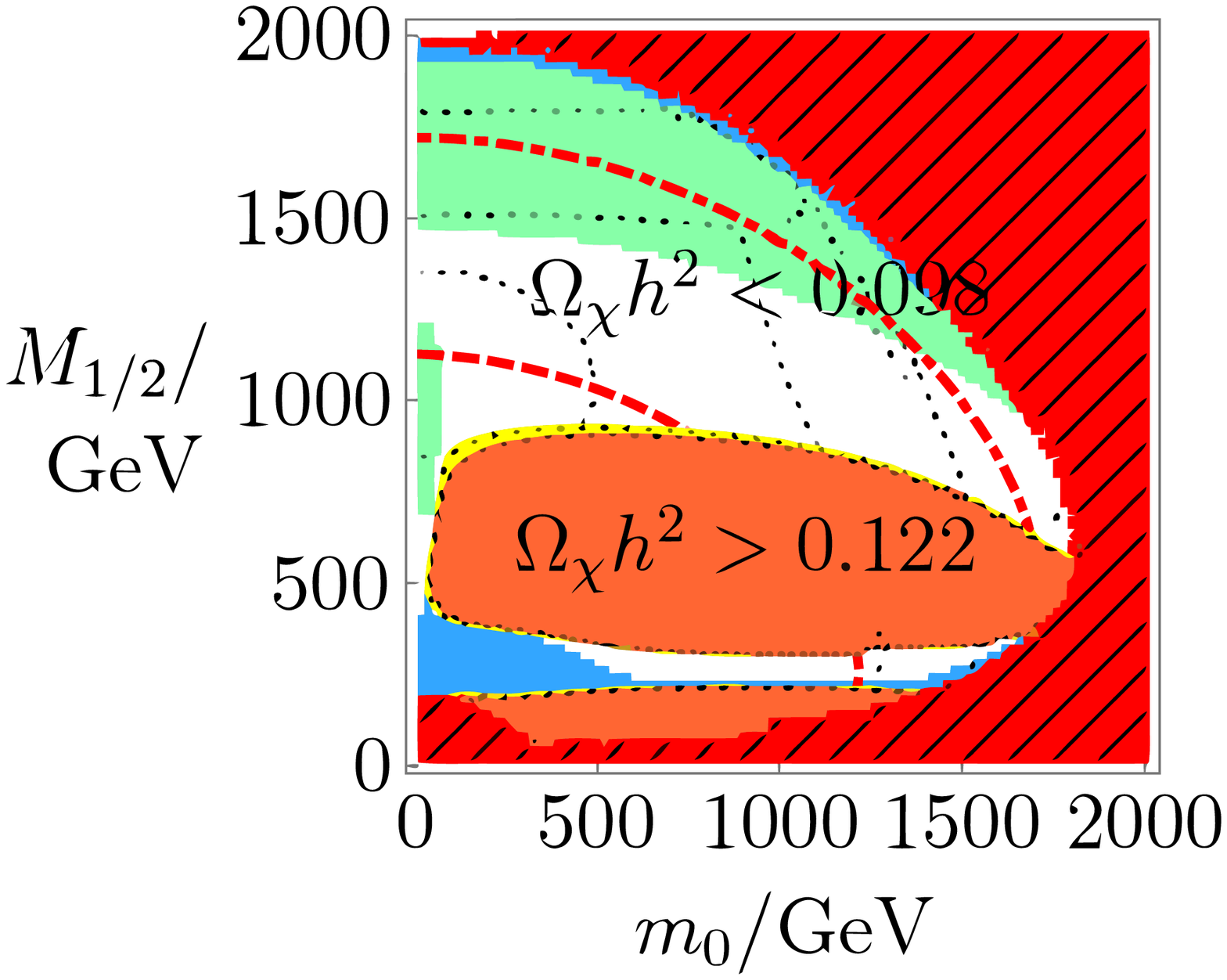}
          }
     \subfigure[210, $\mu<0$]{
          \label{f.210mDM}
          \includegraphics[width=0.45\textwidth]{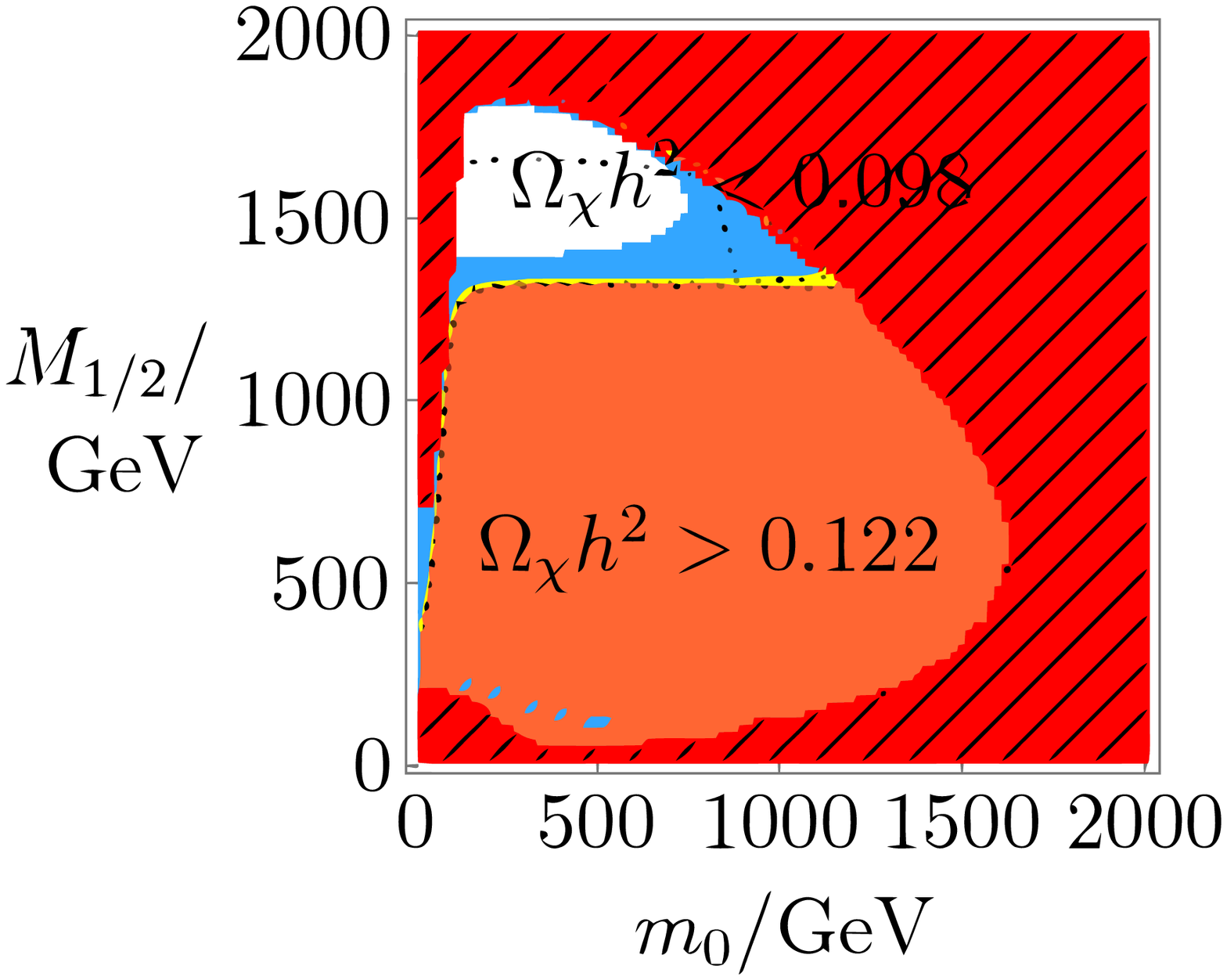}
          }
     \caption{\small Contour plots of $\Omega_{\chi}h^2$ in the 54 and 210 models. These are for the hypersurface in parameter space with $\tan\beta=10$ and  $A_0=0$. The narrow, yellow region, which lies between the regions of over- and under- abundance, has a dark matter abundance that satisfies, within $3\sigma$, the constraint given in Eq~\eqref{e.OmegaCDM}. The (red) dashed and (red) dot-dashed contours indicate where the Higgs mass, $m_{h^0}$, is 111 GeV and 114 GeV, respectively.}
     \label{f.DM}
\end{figure}
In particular, as $M_{1/2}$ increases one expects to pass from a Bino to Higgsino dominated LSP. During this transition one will pass through a so-called `well tempered' region, for which the composition of the neutralino is such that $\langle\sigma v\rangle$ is tuned to give agreement between $\Omega_{\chi}$ and $\Omega_{\rm CDM}$. In order to study this in more detail we have calculated $\Omega_{\chi}h^2$ for the 54 and 210 using micrOmegas \cite{c.micrOmegas}. The results of a scan over the $m_0\text{---}M_{1/2}$ plane are displayed in Fig~\ref{f.DM}, for the parameter choice $A_0=0$ and $\tan\beta=10$. The plot is divided into three main regions, depending upon the neutralino abundance. Within the narrow yellow band the abundance satisfies:
\beq\label{e.OmegaRange}
0.1029\leq\Omega_{\chi}h^2\leq0.1233
\eeq
which agrees with the best fit value given in Eq~\eqref{e.OmegaCDM} to within 3$\sigma$. (Note that we have neglected any errors in the theoretical prediction of the dark matter abundance.) The remaining orange and white regions of the plot indicate an excess or under-abundance of dark matter, respectively. Significantly, one may note that there is a region of the $m_0$---$M_{1/2}$ plane, for both the 54 and 210, in which both $\Omega_{\chi}=\Omega_{\rm CDM}$ and $M_{h^0}\geq 111 \text{ GeV}$ are satisfied, with $\Delta<10$.

There have been several previous studies of neutralino dark matter in the presence of non-universal gaugino masses \cite{c.MHDM0,c.MHDM,c.MHDM1,c.MHDM2}. In particular the case $M_3(M_X)=M_1(M_X)< M_{1/2}$ has previously been considered in \cite{c.MHDM2}, where a well-tempered neutralino, with $\Omega_{\chi}=\Omega_{\rm CDM}$ was shown to be achieved for a suitable value of $M_3/M_{1/2}$. The points in the yellow band of Fig~\ref{f.54DM} are a special case, justified by a GUT, of this more general study, to which we refer the reader for a more detailed discussion of direct and indirect detection, together with collider phenomenology.

The delicate tuning of the parameters required to achieve $\Omega_{\chi}=\Omega_{\rm CDM}$, does suggest that some degree of fine-tuning may be present in this particular `well-tempered' scenario. In order to study this fine-tuning we use the sensitivity measure of Ellis and Olive \cite{c.CDMFT1}, which defines the tuning in $\Omega_{\chi}$ due to a parameter $a_i$ as:
\beq
\Delta^{\Omega}_i=\frac{\partial \ln \Omega_{\chi}}{\partial \ln a_i}\text{ ,}
\eeq
which is analogous to the fine-tuning measure used for EWSB, given in \eqref{e.FTmeasure}. To permit comparison with previous studies of $\Delta^{\Omega}$, performed for different patterns of soft masses, we will utilise the following parameter set:
\beq
a_i\in\{m_0, M_{1/2}, A_0, \tan\beta\}
\eeq
(which differs from that used for fine-tuning in $M_Z$ by treating $\tan\beta$ as a parameter instead of $\mu$) and, following King and Roberts \cite{c.CDMFT3}, define our total fine-tuning $\Delta^{\Omega}={\rm Max}\{|\Delta^{\Omega}_i|\}$.
For the 54 we find that $5 \lesssim \Delta^{\Omega}\lesssim 15$ in the region which has $\Delta<5$ and $M_{h^0}>111 \text{ GeV}$ (increasing as $m_0$ increases), whilst for the 210 $\Delta^{\Omega}\approx 15$ . To put this into context one can consider the fine-tunings observed for other points in MSSM parameter space. Ellis and Olive determined that the CMSSM, with universal gaugino masses, has $\Delta^{\Omega}\lesssim 1$ and $\Delta^{\Omega}\lesssim 5$ in the bulk and stau coannihilation regions, respectively, whilst $\Delta^{\Omega}\approx 200$ near the focus point \cite{c.CDMFT1}. In comparison, a study of non-universal gaugino masses performed by King and Roberts revealed that a well-tempered neutralino, obtained by bino-wino mixing, has $\Delta^{\Omega}\approx 28$ when no relation is assumed between gaugino masses \cite{c.CDMFT3}, which was shown by Birkedal-Hansen and Nelson \cite{c.CDMFT2} to fall to $\Delta^{\Omega}\lesssim  3, 17$ (depending on $\tan\beta$) when one assumes a relation of the form $M_2(M_X)=rM_1(M_X)$ with $r$ taking the fixed value $r=0.6$. Hence, in comparison, the fine-tuning $\Delta^{\Omega}$ present in the well-tempered regions of the 54 and 210, whilst by no means optimal, appears to be fairly mild. But whilst the narrow yellow band seems to satisfy the constraint of Eq~\eqref{e.OmegaRange} with low fine-tunings $\Delta$ and $\Delta^{\Omega}$, it remains impossible, without further model building, to motivate why this small portion of parameter space should be preferred over any other.

\section{Summary and Conclusions}\label{s.Conc}
In this paper we have demonstrated how certain relations between the gaugino masses, present at the scale $M_X$, can lead to a focus point in the soft mass $m_{H_u}^2$ that lies near to the TeV scale. At this focus point $m_{H_u}^2$ has a weak dependence upon the parameter $M_{1/2}$, which we argued in Section~\ref{s.GauginoFP} could allow a large gaugino mass $M_{1/2}\sim1 \text{ TeV}$ with a fine-tuning less than 10. In Section~\ref{s.Models} we demonstrated that these ratios can be generated by known mechanisms of SUSY breaking. In particular, when SUSY breaking is mediated by gravity, we demonstrated that GUT models, through the F-term of some Higgs superfield, can generate ratios that will realise a low scale focus point. We have also demonstrated that this is possible within a class of string model, in which the supersymmetry breaking occurs in the moduli fields. It is particularly interesting, we feel, that the ratios required to realise a focus point can be motivated within the context of GUT and string constructions.

In Section~\ref{s.FT} we computed the fine-tuning of these models numerically, taking into account all one-loop corrections, and demonstrated that several of these models permit a range of parameters for which the Higgs mass is consistent with the LEP bound of 114 GeV, with a fine-tuning less than 5. We would like to stress that the structure of the one-loop corrections is particularly important in these focus point scenarios, as it tends to significantly reduce the levels of fine-tuning expected from a naive tree-level analysis.

Each of these models are characterised by their spectra of gauginos and Higgsinos. In particular one expects a light Higgsino with mass less than 200~GeV, whilst the remaining gauginos and scalars can be significantly heavier, with masses $\gtrsim 1$ TeV. The light Higgsinos in these models can generate significant contributions to ${\rm Br}(\overline{B}\rightarrow X_s\gamma)$. For the case of $A_0=0$ and moderate $\tan\beta$ we have shown that consistency with experiment requires $\mu<0$ in models with $\eta_3>0$, whilst the sign of $\mu$ is weakly constrained in models with $\eta_3<0$. In all of the models considered the region with an acceptable Higgs mass give very small contributions to $a_{\mu}$. Thus the discrepancy $\delta a_{\mu}$ between the measured and predicted values of $a_{\mu}$ is similar to that in the SM. If this result persists, and its statistical significance increases, one will need to introduce further non-universality into the scalar masses to account for this observation. Furthermore we have also identified models which possess a light Bino, thus giving rise to a LSP that is a mixture of Higgsino and Bino states. For some region of parameter space these models possess a so-called `well tempered' LSP, in which $\Omega_{\chi} h^2$ is consistent with the value of $\Omega_{\rm CDM} h^2$ required in the $\Lambda {\rm CDM}$ model. We have examined the degree of tuning present in these well-tempered regions, which we have found to be $\sim \mathcal{O}(10)$. This tuning is relatively mild compared to other scenarios in the MSSM which achieve the correct dark matter abundance.

To conclude, the presence of a focus point in the soft terms of the MSSM may explain why the superpartners are significantly heavier than the weak scale. It is a relatively simple hypothesis, requiring nothing more than the structure of the MSSM up to the scale of $M_X$, that can be justified within known mechanisms of SUSY breaking. Furthermore, by measuring the masses of the supersymmetric states, it is a hypothesis that can be tested in the near future.

\section*{Acknowledgements}
DH is supported by the STFC (PPA/S/S/2005/04179).

\section*{Appendix A: Conventions}
Here we define the conventions that we employ throughout this paper, which are similar to those used within SOFTSUSY \cite{c.SOFTSUSY}. We work within the context of the R-parity conserving MSSM, with the soft SUSY breaking sector given by:
\beq
\mathcal{L_{\rm soft}}= -m_i^2|\phi_i|^2-(M_{\alpha\beta}\lambda^{\alpha}_j\lambda^{\beta}_j+\tfrac{1}{3!}A_{ijk}Y_{ijk}\phi_i\phi_j\phi_k+B\mu H_uH_d+{\rm h.c.})
\eeq
where $\phi_i$ and $\lambda^{\alpha}_j$ are scalar and gaugino fields respectively. The fields $H_u$ and $H_d$ refer to the up and down type Higgs, respectively, and we define $H_uH_d=H_u^0H_d^0-H_u^+H_d^-$. The parameter $\mu$ is the usual Higgs mass that appears in the superpotential. Without any loss of generality we choose our phases such that both $M_2$ and $B\mu$ are real and positive. In order to avoid constraints on CP-violating phases, we further assume that all phases are either $0$ or $\pi$.

\section*{Appendix B: Numerical Solutions of the RGEs}
In this section we summarise some of the RG coefficients, defined in Eq~\eqref{e.RGEsolnsSOFT}, which are used in Section~\ref{s.GauginoFP}.

These coefficients are solutions to the two loop RGEs \cite{c.RGEs}. They are obtained using the boundary conditions for the Snowmass point SPS 1a, which was obtained using SOFTSUSY \cite{c.SOFTSUSY}. The relevant quantities that define the boundary conditions are as follows:
\begin{align}
M_X&=2.38\times 10^{16}\nonumber\\
g_1(M_X)&=g_2(M_X)=0.721\nonumber\\
g_3(M_X)&=0.708\nonumber \\
h_t(M_X)&=0.494\nonumber \\
h_b(M_X)&=0.050\nonumber \\
h_{\tau}(M_X)&=0.069
\end{align}

The relevant coefficients are given in the Tables~\ref{t.Scoeffs} and \ref{t.Gcoeffs} below, evaluated at a renormalization scale $Q=500\text{ GeV}$. Note that each row contains the contribution of each gaugino ratio, or combination of ratios, to a coefficient. In our conventions $\eta_2=1$.

\begin{table}[h]
\begin{center}
\begin{tabular}{|c|c|c|c|c|c|c|}
\hline
 &$\eta_3^2$ &$\eta_3\eta_2$&$\eta_3\eta_1$&$\eta_2^2$&$\eta_1\eta_2$&$\eta_1^2$\\
\hline
$z_{H_d}^M$ & -0.051 & -0.018 & -0.001 & 0.438 & -0.001 & 0.036\\
$z_{H_u}^M$ & -1.84 & -0.157 & -0.023 & 0.211 & -0.006 & 0.006 \\
$z_{Q3}^M$ & 3.671 & -0.100 & -0.011 & 0.345 & -0.002 & -0.006\\
$z_{U3}^M$ & 3.172 & -0.119 & -0.020 & -0.158 & -0.003 & 0.042 \\
\hline
\end{tabular}
\end{center}
\caption{\small The coefficients $z_i^M$, evaluated at a renormalization scale $Q=500\text{ GeV}$ for the high scale boundary conditions appropriate for SPS 1a.}
\label{t.Scoeffs}
\end{table}

\begin{table}[h]
\begin{center}
\begin{tabular}{|c|c|c|c|}
\hline
 &$\eta_3$ &$\eta_2$&$\eta_1$\\
\hline
$\mathcal{Z}_{1}^M$ & -0.010 & 0 & 0.418 \\
$\mathcal{Z}_{2}^M$ & -0.028 &0.794 & 0 \\
$\mathcal{Z}_{3}^M$ & 2.350 & -0.010 & -0.001 \\
$\mathcal{Z}_{A_t}^M$ & -1.607 & -0.249 & -0.033 \\
\hline
\end{tabular}
\end{center}
\caption{\small The coefficients $\mathcal{Z}_i^M$, evaluated at a renormalization scale $Q=500\text{ GeV}$ for the high scale boundary conditions appropriate for SPS 1a.}
\label{t.Gcoeffs}
\end{table}


\begin{thebibliography}{99}
\bibitem{c.LowEnergySUSY} H. P. Nilles, Physics Reports 110 (1984) 1.

\bibitem{c.MSSM} S. Dimopoulos and H. Georgi, Nucl. Phys. B 193 (1981) 150.

\bibitem{c.LEPHiggsLimit} ALEPH, DELPHI, L3 and OPAL Collaborations, The LEP Working Group for Higgs Boson Searches, Phys. Lett. B 565 (2003) 61, hep-ex/0306033;\\
    LEP Higgs Working Group, Eur. Phys. J. C 47 (2006) 547, hep-ex/0602042.

\bibitem{c.LEPCharginoMassLimits} ALEPH Collaboration, Phys. Lett. B 533 (2002) 223, hep-ex/0203020.

\bibitem{c.FTPriceLEP.CEP} P.H. Chankowski, J.R. Ellis, S. Pokorski, Phys. Lett. B 423 (1998) 327, hep-ph/9712234.
\bibitem{c.FTPriceLEP.BS} R. Barbieri and A. Strumia, Phys. Lett. B 433 (1998) 63, hep-ph/9801353.
\bibitem{c.FTPriceLEP.CEOP} P.H. Chankowski, J.R. Ellis, M. Olechowski, S. Pokorski, Nucl. Phys. B 544 (1999) 39, hep-ph/9808275.
\bibitem{c.FTPriceLEP.KK} G.L. Kane and S.F. King, Phys. Lett. B 451 (1999) 113, hep-ph/9810374.


\bibitem{c.EDMs} T. Ibrahim and P. Nath, Rev. Mod. Phys. 80 (2008) 577, arXiv:0705.2008 [hep-ph].

\bibitem{c.Essig} R. Essig, Phys. Rev. D 75 (2007) 095005, hep-ph/0702104.

\bibitem{c.DermiGunion} R. Derm\'{i}\~{s}ek and J.F. Gunion, Phys. Rev. D 77 (2008) 015013, arXiv:0709.2269 [hep-ph].

\bibitem{c.IR} L.E. Ib\'{a}\~{n}ez and G.G. Ross, Phys. Lett. B 110 (1982) 215.

\bibitem{c.BarbGiud} R. Barbieri and G. Giudice, Nucl. Phys. B 306 (1988) 63.


\bibitem{c.Hyperbolic} K.~L.~Chan, U.~Chattopadhyay and P.~Nath, Phys. Rev. D 58 (1998) 096004, hep-ph/9710473

\bibitem{c.Focus} J.L. Feng, K.T. Matchev and T. Moroi, Phys. Rev. D 61 (2000) 075005, hep-ph/9909334.

\bibitem{c.NaturalnessfromStringRelations} G. L. Kane, J. Lykken, B. D. Nelson and L. Wang, Phys. Lett. B 551 (2003) 146, hep-ph/0207168.

\bibitem{c.ModAnom} K. Choi, K.S. Jeong, T. Kobayashi and K. Okumura, Phys. Rev. D 75 (2007) 095012, hep-ph/0612258.

\bibitem{c.CW} S.R. Coleman and E.J. Weinberg, Phys. Rev. D 7 (1973) 1888.

\bibitem{c.AKO} H. Abe, T. Kobayashi, Y. Omura, Phys. Rev. D 76 (2007) 015002, hep-ph/0703044.

\bibitem{c.RGEs} S.P. Martin and M.T. Vaughn, Phys. Rev. D 50 (1994) 2282, hep-ph/9311340 [Erratum: {\it ibid} D78 (2008) 039903].

\bibitem{c.SOFTSUSY} B.C. Allanach, Comput. Phys. Commun. 143 (2002) 305, hep-ph/0104145.

\bibitem{c.PDG} W.-M. Yao et al., J. Phys. G 33 (2006) 1.
\bibitem{c.TopQuarkMass} Tevatron Electroweak Working Group, arXiv:0803.1683 [hep-ex].

\bibitem{c.SPS} B.C. Allanach {\it et. al.}, Eur. Phys. J. C 25 (2002) 113, hep-ph/0202233.

\bibitem{c.Cremmer} E. Cremmer, S. Ferrara, L. Girardello and A. Van Proeyen, Phys. Lett. B 116 (1982) 231.

\bibitem{c.NonUniGauginos} J.R. Ellis, K. Enqvist, D.V. Nanopoulos and K. Tamvakis, Phys. Lett. B 155 (1985) 381.
\bibitem{c.SO10GauginoMasses} N. Chamoun, C. Huang, C. Liu and X. Wu, Nucl. Phys. B 624 (2002) 81, hep-ph/0110332.
\bibitem{c.MartinNonUni} S.P. Martin, Phys. Rev. D 79 (2009) 095019, arXiv:0903.3568 [hep-ph].

\bibitem{c.TheoryOfSoftTerms} A. Brignole, L.E. Ib\'{a}\~{n}ez and C. Mu\~{n}oz, Nucl. Phys. B422 (1994) 125, hep-ph/9308271 [Erratum: {\it ibid.} B436 (1995) 747].
\bibitem{c.StringSoft} P. Bin\'{e}truy, M. K. Gaillard and B. D. Nelson, Nucl. Phys. B 604 (2001) 32, hep-ph/0011081.

\bibitem{c.MMA} K. Choi, K.S. Jeong, T. Kobayashi, K. Okumura, Phys. Lett. B 633 (2006) 355, hep-ph/0508029.

\bibitem{c.HiggsUncertainty} B.C. Allanach, A. Djouadi, J.L. Kneur, W. Porod, P. Slavich, JHEP 0409 (2004) 044, hep-ph/0406166.

\bibitem{c.CHAMPconstraints} M. Taoso, G. Bertone and A. Masiero, JCAP 0803 (2008) 022, arXiv:0711.4996 [astro-ph].

\bibitem{c.RomaninoStrumia} A. Romanino and A. Strumia, Phy. Lett. B 487 (2000) 165, hep-ph/9912301.

\bibitem{c.degenrateHiggsinos} G.F. Giudice, A. Pomerol, Phys. Lett. B 372 (1996) 253, hep-ph/9512337; D. Pierce, A. Papadopoulos, Phys. Rev. D 50 (1994) 565, hep-ph/9312248.

\bibitem{c.LightNeutralinos} H.K. Dreiner, S. Heinemeyer, O. Kittel, U. Langenfeld, A.M. Weber and G. Weiglein, Eur. Phys. J. C 62 (2009) 547, arXiv:0901.3485 [hep-ph].

\bibitem{c.HFAG} Heavy Flavour Averaging Group, arXiv:0808.1297 [hep-ex].

\bibitem{c.bsgSM} M. Misiak et. al., Phys. Rev. Lett. 98 (2007) 022002, hep-ph/0609232.

\bibitem{c.superISO} F. Mahmoudi, Comput. Phys. Commun. 178 (2008) 745, arXiv:0710.2067 [hep-ph];\\ F. Mahmoudi, arXiv:0808.3144 [hep-ph].

\bibitem{c.SignBSG} R. Garisto and J.N. Ng, Phys. Lett. B 315 (1993), 372, hep-ph/9307301.

\bibitem{c.gMinus2Result} G. W. Bennet {\it et. al.}, Phys. Rev. D 73 (2006) 072003, hep-ex/0602035.
\bibitem{c.gMinus2Pred} K. Hagiwara, A.D. Martin, D. Nomura and T. Teubner, Phys. Lett. B 649 (2007) 173, hep-ph/0611102.
\bibitem{c.aMUdiscrep} J.P. Miller, E. de Rafael and B.L. Roberts, Rept. Prog. Phys. 70 (2007) 795, hep-ph/0703049.


\bibitem{c.WMAP5yr1} WMAP Collaboration (J. Dunkley {\it et. al.}), Astrophys. J. Suppl. 180 (2009) 306, arXiv:0803.0586 [astro-ph].
\bibitem{c.WMAP5yr2} WMAP Collaboration (E. Komatsu {\it et. al.}), Astrophys. J. Suppl. 180 (2009) 330, arXiv:0803.0547 [astro-ph].

\bibitem{c.MHDM0} V. Bertin, E. Nezri and J. Orloff, JHEP 0302 (2003) 046, hep-ph/0210034.
\bibitem{c.MHDM} G. Bélanger, F. Boudjema, A. Cottrant, A. Pukhov and A. Semenov, Nucl. Phys. B 706 (2005) 411, hep-ph/0407218.
\bibitem{c.MHDM1} H. Baer, A. Mustafayev, S. Profumo and X. Tata, Phys. Rev. D 75 (2007) 035004, hep-ph/0610154.
\bibitem{c.MHDM2} H. Baer, A. Mustafayev, H. Summy and X. Tata, JHEP 0710 (2007) 088, arXiv:0708.4003 [hep-ph].

\bibitem{c.EarlyUni} E.W. Kolb and M.S. Turner, The Early Universe, Perseus Publishing, 1994.

\bibitem{c.micrOmegas} G. Bélanger, F. Boudjema, A. Pukhov and A. Semenov, Comput. Phys. Commun. 174 (2006) 577, hep-ph/0405253;\\ G. Bélanger, F. Boudjema, A. Pukhov and A. Semenov, Comput. Phys. Commun. 149 (2002) 103, hep-ph/0112278.

\bibitem{c.WellTempNeut}  N. Arkani-Hamed, A. Delgado, G.F. Giudice, Nucl. Phys. B 741 (2006) 108, hep-ph/0601041.

\bibitem{c.HiggsinDM} V. Bertin, E. Nezri and J. Orloff, JHEP 0302 (2003) 046, hep-ph/0210034;\\ A. Birkedal-Hansen and B.D. Nelson, Phys. Rev. D 67 (2003) 095006, hep-ph/0211071;\\ D.G. Cerde\~{n}o and C. Mu\~{n}oz, JHEP 0410 (2004) 015, hep-ph/0405057;\\ U. Chattopadhyay and D.P. Roy, Phys. Rev. D 68 (2003) 033010, hep-ph/0304108.



\bibitem{c.CDMFT1} J.R. Ellis, K.A. Olive, Phys. Lett. B 514 (2001) 114, hep-ph/0105004.
\bibitem{c.CDMFT2}  A. Birkedal-Hansen and B.D. Nelson, Phys. Rev. D 67 (2003) 095006, hep-ph/0211071.
\bibitem{c.CDMFT3} S.F. King and J.P. Roberts, JHEP 0609 (2006) 036, hep-ph/0603095;\\ S.F. King, J.P. Roberts and D.P. Roy, JHEP 0710 (2007) 106, arXiv:0705.4219 [hep-ph].


\end{thebibliography}
\end{document}